\documentclass{aa}
\usepackage{graphicx}

\normalfont

\newcommand{\lsim}{\raise0.3ex\hbox{$<$}\kern-0.75em{\lower0.65ex\hbox{$\sim$}}}
\newcommand{\gsim}{\raise0.3ex\hbox{$>$}\kern-0.75em{\lower0.65ex\hbox{$\sim$}}}

\newcommand{\cmc}{\rm ~cm^{-3}}

\newcommand{\kms}{\rm ~km~s^{-1}}

\newcommand{\wl}{\lambda}

\def\EE#1{\times 10^{#1}}

\begin{document}         

\title{X-ray emission from radiative shocks in Type~II supernovae}
\author{Tanja K. Nymark
\and Claes Fransson
\and Cecilia Kozma}
\titlerunning{Circumstellar Interaction}
\authorrunning{Nymark et al.}
\institute{Stockholm Observatory, AlbaNova University Center, SE-106 91 Stockholm, Sweden}
\offprints{Tanja K. Nymark, \email{tanja@astro.su.se}}

\date{Received  8 Sep 2005/ Accepted 21 Oct 2005 }

\abstract{The X-ray emission from the circumstellar interaction in
  Type~II supernovae with a dense circumstellar medium is
  calculated. In Type~IIL and Type~IIn supernovae mass loss rates are
  generally high enough for the region behind the reverse shock to be
  radiative, producing strong radiation, particularly in X-rays. We
  present a model for the emission from the cooling region in the case
  of a radiative reverse shock. Under the assumption of a stationary
  flow, a hydrodynamic model is combined with time dependent
  ionization balance and multilevel calculations. The applicability of
  the steady state approximation is discussed for various values of
  the ejecta density gradient and different sets of chemical
  composition. We show how the emerging spectrum depends strongly on
  the reverse shock velocity and the composition of the shocked
  gas. We discuss differences between a spectrum produced by this
  model and a single-temperature spectrum. Large differences for
  especially the line emission are found, which seriously can affect
  abundance estimates. We also illustrate the effects of absorption in
  the cool shocked ejecta. The applicability of our model for various
  types of supernovae is discussed.

}
\maketitle

\section{Introduction}
\label{sect:intro}
\begin{figure}[h]
\begin{center}
\resizebox{\hsize}{!}{\includegraphics{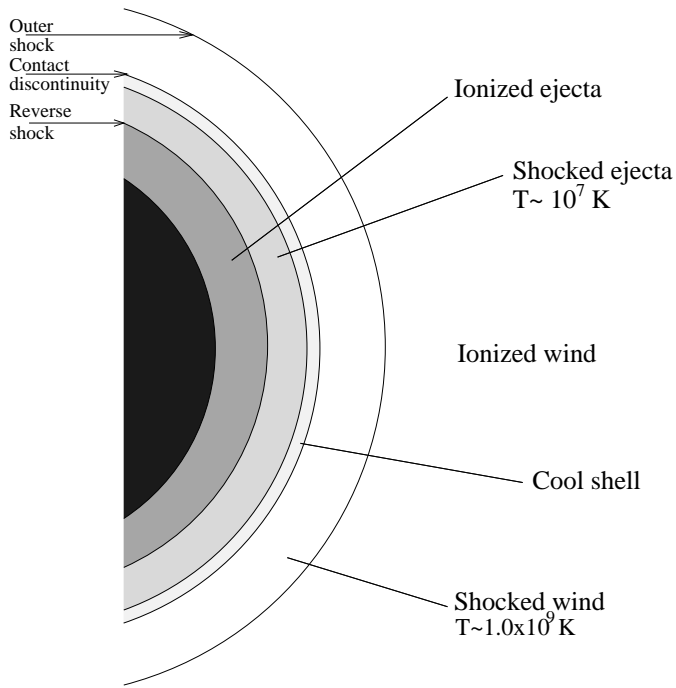}}
\end{center}
\caption{Schematic picture of the different regions in the interaction of supernova ejecta with the progenitor wind. The figure is not to scale. The contact discontinuity between the shocked wind and shocked ejecta is Rayleigh-Taylor unstable.} 
\label{fig:shock_structure}
\end{figure}

The emission at late times is for most Type~IIL (Type II linear) and
Type~IIn (Type II narrow line) supernovae dominated by the interaction between the supernova ejecta and the circumstellar medium. The collision between the two creates two shocks; one moving outward into the circumstellar medium and a reverse shock, which moves backward into the ejecta (Fig.~\ref{fig:shock_structure}). Because of the high temperatures involved, most of the emission from the interaction region is radiated as X-rays. The strong emission heats and ionizes the circumstellar medium and ejecta, leading to line emission in the optical and UV (Fransson \cite{Fransson1984}; Chevalier \& Fransson 1994, hereafter \cite{CF94}). This can then act as probes of the ejecta and the shock. Another characteristic feature of the interaction is strong radio emission caused by synchrotron emission behind the outer shock (e.g., Sramek et al.~\cite{Sramek03}).
A review of circumstellar interaction in supernovae
is given by Chevalier  \& Fransson (\cite{CF03}). 

Chevalier (\cite{Chevalier1982a}, \cite{Chevalier1982b}) proposed circumstellar
interaction as an explanation of the observed radio light curves of
SN~1979C and SN~1980K, and used a self-similar formulation to analyze
the effects of interaction between supernova ejecta and the ambient
medium. Different mechanisms for production of X-rays were also discussed. In Fransson (\cite{Fransson1984}) the X-ray emission from the reverse and circumstellar shocks was discussed in some detail. In particular, it was argued that in most cases the reverse shock would be radiative, a fact we will exploit in this paper. This also has the consequence that a cool, dense shell will form between the two shocks, which is important for both the X-ray and optical emission. A detailed study of the effects of
circumstellar interaction in Type II supernovae was done by Chevalier
\& Fransson (\cite{CF94}), while the specific case of
SN~1993J was investigated by Fransson et al.~(1996, hereafter \cite{FLC96}), as well as by Suzuki \& Nomoto (\cite{SuzNom95}). 

Radiative shocks in supernova remnants were first modeled by Cox (\cite{Cox72}). Later, improved models were made by Raymond (\cite{Ray79}), Dopita (\cite{Dopita76}, \cite{Dopita77}) and others. Because of the lower densities, the shocks are, however, only radiative for velocities $\la 200\ \kms$, substantially lower than the velocities studied in the present paper. High density models, suitable for active galactic nuclei (AGN) and young supernova remnants were made by  Plewa \& $\mathrm{R\acute{o}\dot{z}yczka}$~(\cite{Plewa92}) and Plewa~(\cite{Plewa93}, \cite{Plewa95}). These models are, however, mainly concerned with the hydrodynamics, with a simplified treatment of the emergent spectra. 

The goal of the work in this paper has therefore been to develop a
model for the X-ray emission from the reverse shock. This combines
the hydrodynamic structure with a spectral code which, by including
updated atomic data and multi-level solutions to most ions, can model
the line emission in greater detail than has been done before.

In Sect.~\ref{sect:obs} we give a brief summary of the most important
observations of circumstellar interaction, and in particular X-ray
observations. Our model is described in Sects.~\ref{sect:hydro} through
\ref{sect:emission}, while Sect.~\ref{sect:discussion} describes the
results of our calculations, including the contribution from the
circumstellar shock. The conclusions are summarized in Sect.~\ref{sect:concl}.

\section{X-ray emission from supernovae}
\label{sect:obs}
The main observational indications of circumstellar interaction are
the detection of narrow optical and UV emission lines from ionized circumstellar gas,
broad, box-like emission lines from ejecta ionized by radiation from the
interaction region, and a wavelength dependent turn-on of the observed
radio emission. The latter is explained as a result of either
free-free absorption or synchrotron self-absorption. Another effect of
circumstellar interaction is that the supernova at late epochs displays a
flat optical light curve, long after the emission caused by
radioactive decay has faded away. For Type~IIn supernovae circumstellar interaction dominates the optical spectrum already from early epochs. We refer to Filippenko (\cite{Filippenko97}) for a review of the optical observations. 

Besides the radio emission, the X-rays show the clearest evidence for circumstellar interaction. Because the X-ray emission is the focus of this paper, we will briefly review the observational evidence for this.

Until now, $\sim 24$ supernovae have been detected in X-rays, and 
for $\sim 10$ of these the data are of sufficiently good quality to be studied in detail. We will concentrate on those for which good data exist, and circumstellar interaction is thought to be important. 
Comprehensive reviews of the existing X-ray observations of supernovae can be found in Schlegel (\cite{Schlegel95}) and Immler \& Lewin (\cite{ImmLew02})\footnote{A complete list of the X-ray supernovae may be found at http://lheawww.gsfc.nasa.gov/users/immler/supernovae\_list.html}.

The first supernova to be detected in X-rays
was SN~1980K, which was observed by the Einstein satellite 35 days
after outburst (Canizares et al.~\cite{Canizares82}). At the time of the next observation at 82 days, the emission had decayed by at least a factor of two. The emission was found to be rather soft, with $kT\sim 0.5$ keV, but no spectra were obtained. Also SN~1979C was observed with Einstein at early epochs, but no source was seen (Panagia et al.~\cite{Panagia80}). After $\sim 16$ years it was, however, detected as a strong source by ROSAT (Immler et al.~\cite{Immler98}). It was observed by the
ASCA satellite  in 1997 (Ray et al.~\cite{Ray01}) and by Chandra
in 1999 and 2000 (Kaaret~\cite{Kaaret01}), with a
roughly constant X-ray flux in the period of 16--20 years after
outburst. The emission was found to be soft, and likely to originate in the shocked ejecta. The reason for the absence of strong X-rays at early epochs is most likely absorption by the cool, dense shell formed by the reverse shock (\cite{CF94}).

The Type IIn SN~1986J in NGC~891 was observed with ROSAT and ASCA from
nine to thirteen years after outburst (Houck et
al.~\cite{Houck98}). During this period it was found to have a fast decline of
the X-ray emission, with $L_{x}\propto t^{-2}$. From the ASCA observations it was found to have a rather
hard spectrum ($kT\sim 5-7$ keV) with a clear Fe~K emission line at 6.7 keV, with a width $< 20\,000\ \mathrm{km\ s^{-1}}$.

The first supernova to be identified through its X-ray emission was
the Type IIn SN~1978K. Strong
radio and optical emission had been detected earlier (Ryder~\cite{Ryder93};
Dopita \& Ryder~\cite{Dopita90}), but it was with the detection of X-rays by
ROSAT in 1992 (Ryder~\cite{Ryder93}) that it was identified as a
supernova. Extensive observations in radio, optical, UV and X-rays were carried out in the late 1990s, and observations by Newton-XMM in October 2000, and by Chandra two years later, indicated a constant X-ray
flux over the past decade (Schlegel et al.~\cite{Schlegel04}). In the early X-ray spectra no emission lines were found, but in the Chandra data lines from Si are evident.                  

The best observed case of circumstellar interaction is SN~1993J in M81. Since its discovery in March 1993, it has been extensively studied in all wavebands. SN~1993J showed signs of
circumstellar interaction early on, first through the detection of
radio emission on day 5 (Weiler et al.~\cite{Weiler93}), and shortly thereafter
through X-ray observations. The first X-ray detection was made by
ROSAT as early as 6 days after discovery (Zimmerman et al.~\cite{Zim94}), and the first observation by ASCA came two days later. The early ASCA observations showed a strong Fe~K line at 6.7~keV, with a width of $0.3\pm 0.3$ keV, which faded within weeks after the explosion (Kohmura et al.~\cite{Koh94}, Uno et al.~\cite{Uno02}). Both ROSAT and ASCA observations showed a hard spectrum during the first two months. This was confirmed by OSSE observations on Compton/GRO,
where hard X-ray emission with $kT\sim 50-150$~keV, was detected in two observations,
9.9--15.4 days after explosion and again between days 23.7--36.9 (Leising et al.~\cite{Leising94}). Three
months after explosion it was too weak to be observed by OSSE.
When it was re-observed by ASCA and ROSAT after six months a soft component was found to dominate the spectrum. 
This transition from a hard to a soft spectrum was explained by \cite{FLC96} as a result of decreasing optical depth in the cool shell between the two shocks, which allows the soft X-rays from the reverse shock to escape.

Very late Chandra observations on day 2594 yielded a detailed spectrum of
SN~1993J in the 0.3-8.0~keV band (Swartz et al.~\cite{Swartz03}). This spectrum showed evidence of
Fe~L emission, as well as emission from H-like Mg and Si and He-like
Mg, Si and S. This is taken by Swartz et al. to indicate the presence of two
low-temperature components, while the high energy ($\gsim
1.5$ keV) continuum emission is thought to point to a high-temperature
component as well.

The Type~IIn SN~1995N was
detected by ROSAT in 1996. The explosion date is unknown, but Fransson et al.~(\cite{Fransson2002}) argue that based on the spectrum a likely explosion date was July 1994. It was observed by ROSAT in 1996 and 1997, by ASCA in 1998
(Fox et al.~\cite{Fox00}) and by Chandra in 2004 (Chandra et al.~\cite{Chandra05}). The X-ray luminosity was found to decline between
the first two observations. Fox et al.~(\cite{Fox00}) found that the luminosity increased again by a factor of two from 1997 to 1998, although a reanalysis by Chandra et al.~(\cite{Chandra05}) of the ASCA observations indicate that the source was contaminated and that both the ASCA and Chandra observations are consistent with a linear decline of the X-ray luminosity. Chandra et al.~(\cite{Chandra05}) also find clear evidence of Ne line emission around 1~keV.

The other well observed Type~IIn SN~1998S in NGC~3877 was discovered at an early phase,
and has been well studied at all wavelengths. It was observed by
Chandra on five occasions from day 678 to day 1324 (Pooley et
al.~\cite{Pooley02}).  A number of emission lines from metals were
observed at energies above 1~keV, among them lines from Ne, Si, S and
Fe. The strengths of the observed lines were claimed to indicate a
high overabundance of the metals with respect to solar values. This
would imply that the observed X-ray emission mainly comes from the
shocked oxygen core. In this case the heavy elements must be mixed out
to a high velocity, possibly caused by an aspherical explosion (Pooley
et al.~\cite{Pooley02}). The abundance analysis is, however, based on
a one-temperature model of the spectrum. In Sect.~\ref{sect:comp1T} we
show that this assumption is highly questionable.

SN~1999em was the first Type IIP supernova to be detected both in X-rays and radio. It was detected by Chandra three days after discovery, and was
observed seven times in a period of almost two years. During this period the spectrum softened from a $\sim 5$~keV spectrum to a $\sim1$~keV spectrum. Pooley et
al. (\cite{Pooley02}) modeled the X-ray emission as resulting from interaction
between supernova ejecta and the wind of a red supergiant, and found that the X-ray
emission agrees well with this, and that both shocks were adiabatic.

The most recent X-ray supernova is the Type~IIP SN~2004dj, which was detected by Chandra in August 2004, most likely two months after explosion. It was found to have a temperature of $kT\sim$7~keV, and nearly three times the luminosity of SN~1999em (Pooley \& Lewin~\cite{PoolLew04}). Modeling of Type~IIP supernovae is discussed in detail by Chevalier et al.~(\cite{CFN05}).

\section{Hydrodynamics}
\label{sect:hydro}

\subsection{Progenitor wind}
The immediate surroundings of the exploding star consist of the
stellar wind of the progenitor. The density of a wind with constant mass loss rate and velocity drops
as $r^{-2}$, but the wind parameters depend on the
evolutionary state of the star. On the main sequence the star is a
blue supergiant with a fast wind. It then evolves to become a red
supergiant with a correspondingly slow and dense wind, with mass loss
rate of $\dot{M} \sim 10^{-6}-10^{-4}~\mathrm{ M_{\odot}\ yr^{-1}}$
and a velocity of $v_{\mathrm{w}}\sim 5-50\ \mathrm{km\ s^{-1}}$ (e.g., Salasnich
et al.~\cite{Salasnich99}). Most massive stars explode in this stage
as supernovae. Stars more massive than $\sim 22\ M_{\odot}$ (Meynet \&
Maeder~\cite{MeynetMaeder2003}) may, however, evolve back to the blue and become Wolf-Rayet stars or, in some cases, end their lives as blue supergiants. 

As the star evolves, the wind will create
an onion-like structure of circumstellar gas, with dense shells
alternating with low density shells. In some stages the progenitor may undergo 
pulsations~(Heger et al.~\cite{Heger97}), which could result in a wind with a density differing from $\rho
\propto r^{-2}$. The circumstellar
medium can also be modified by mass loss in a binary system, as might
have been the case for SN~1993J (e.g., Suzuki~\cite{Suz94}; Suzuki \& Nomoto~\cite{SuzNom95}; Nomoto et
al.~\cite{Nomoto95}). Because of this, the assumption
that the density drops as distance squared may not always hold. A
more general description of the wind density is therefore

\begin{equation}
\label{eq:windDens}
\rho _{\mathrm{w}}=\frac{\dot{M} }{4\pi R_{*}^{2}v_{\mathrm{w}}}\left 
(\frac{R_{*}}{r}\right )^s, 
\end{equation}

\noindent
where $R_*$ is a reference radius. For $s=2$ this
reduces to the expression for a steady wind. There are few cases for deviations from $s=2$ winds (see Fransson \& Bj\"ornsson, \cite{FB05} for a discussion). In some cases, the supernova may, however, interact with either clumpy winds or shell-like structures, for which $s$ may differ from 2 (see Sect.~\ref{sect:intReg}).

\subsection{Ejecta}
The unshocked supernova ejecta are assumed to expand freely, so that the
position of a specific mass element can be expressed as $R(t)=Vt$, where
the velocity $V$ is constant. The structure of the ejecta reflects the
structure of the progenitor  and the explosion
itself. 

A useful approximation to the density
structure of the outer ejecta is

\begin{equation}
\label{eq:ejStruct}
\rho_{\mathrm{ej}}=\rho\left (\frac{t}{t_0}\right )^{-3}\left (\frac{V_0t}{r}\right )^{\eta},
\end{equation}

\noindent
where $\eta $ is the power law index characterizing the outer ejecta.
Analytical ejecta models by Matzner \& McKee (\cite{MatznerMcKee99})
indicate that $\eta =11.9$ for a red
supergiant progenitor and $\eta =10.1$ for a blue supergiant
progenitor. For this to be valid, the progenitor structure has to be close to a polytrope
of index 3/2 (convective envelope) or 3 (radiative envelope), respectively, which does not always hold. In particular, the models
of Matzner \& McKee do not apply to superadiabatic regions, which can be present in certain
progenitors. This may have been the case for SN~1993J (Nomoto et
al.~\cite{Nomoto95}). 

In more detailed hydrodynamical explosion models the power law index $\eta $ is usually in the range 7--12, but can be as
high as $\sim $~20, as is the case in the ejecta models of SN~1993J. Evolutionary progenitor models (Shigeyama \& Nomoto~\cite{ShiNom90};
Woosley \& Weaver~\cite{WooWea95}) indicate that a power law structure is a
good approximation for the outer layers, and to some extent also in the
inner parts, albeit with a different value for the power law index,
$\eta $. Roughly speaking, most  one dimensional models show a
steep inner power law, a shallow middle part and a steep outer part,
although there is considerable structure, particularly in the
inner parts. However, hydrodynamic instabilities at especially the H/He interface as well as the He/O core can modify the structure appreciably (MÂ\"uller et al. \cite{MFA91}; Iwamoto et al. \cite{Iwamoto97}; Kifonidis et al. \cite{Kifonidis03}). 

During the first weeks or months, the reverse shock passes through
the outermost steep part and enters the shallower gradient toward the middle.
The shallow part is traversed quickly, and eventually the shock enters 
the inner steep part. As the shock moves into a region with a different density
slope, the X-ray spectrum will change. The time at which this spectral
change occurs, could help constrain the explosion models. We will discuss this in more detail in a later paper.

\subsection{Interaction region}
\label{sect:intReg}
Chevalier (\cite{Chevalier1982a}) found a  self-similar solution for the interaction
between supernova ejecta and a stationary medium, which provided a complete solution to the physical quantities of the two shocks (see also Nadyozhin~\cite{Nad85}). The main
parameters are the power law exponents, $\eta $ and $s$, of the
ejecta and wind densities, the mass loss parameter $\dot{M}/v_{w}$ and the shock velocity (or equivalently the total energy). 

These relations can be simplified further if one
assumes that the interaction shell is very thin compared to its radius
(the thin shell approximation), which is usually an adequate approximation. This becomes exact in the limit of strong radiative cooling, which is the case for the reverse shocks discussed in this paper. The circumstellar shock is, however, in general adiabatic. A similarity solution describing this case is discussed in \cite{CF94}. In the following the subscript `0' refers to quantities immediately behind the reverse shock, `rev' refers to parameters of the shock itself, while the subscript `cs' refers to the circumstellar shock. 

The density and temperature behind the reverse shock can in the thin shell approximation be written as (\cite{FLC96})

\begin{equation}
\label{eq:densRev}
\rho _0=\frac{(\eta -3)(\eta -4)}{(3-s)(4-s)}\rho _{\mathrm{cs}}
\end{equation}

\begin{equation}
\label{eq:Trev}
T_0=\frac{\mu }{\mu _{\mathrm{cs}}}\frac{(3-s)^2}{(\eta -3)^2}T_{\mathrm{cs}} =2.27\times10^9 {\mu }\frac{(3-s)^2}{(\eta -s)^2}V_{4}^{2}\ \ \ \mathrm{K} 
 \end{equation}

\noindent
where $\mu $ is the mean mass per particle (i.e., electrons plus ions) behind the reverse shock, and $\mu _{\mathrm{cs}}$ is the corresponding value behind the circumstellar shock. 
\noindent
The density behind the outer shock is related
to the wind density by $\rho _{\mathrm{cs}}=4\rho _{\mathrm{w}}$.  Setting $r=R_{\mathrm{s}}=V_{\mathrm{ej}}t$ in Eq.~(\ref{eq:windDens}),  we can express the post-shock ion number density as 

 \begin{equation}
\label{eq:densRev2}
n_0=1.2\times 10^{38}\frac{(\eta -3)(\eta -4)}{(3-s)(4-s)}\frac{\tilde{A}_{*}}{\mu _{\mathrm{A}} V_{\mathrm{ej}}^{s}t^{s}}R_{*}^{s-2}\ \ \ \ \ \ \mathrm{cm}^{-3},
\end{equation}

\noindent
where $\mu _{\mathrm{A}}$ is the mean atomic weight, and we have defined $\tilde{A}_{*}=\dot{M}_{-5}/v_{\mathrm{w}_{1}}=(\dot{M}/10^{-5}M_{\odot})/(v_{\mathrm{w}}/10 \kms )$. Note that $\tilde{A}_{*}=100A_{*}$, where $A_{*}$ was defined by Chevalier \& Li~(\cite{ChevLi99}) for parameters typical of a Wolf-Rayet wind. $R_{\mathrm{s}}$ is the radius of the interaction region, which we here take to be the same as the position of the reverse shock, and $V_{\mathrm{ej}}$ is the maximum velocity of the ejecta, i.e., the velocity of the ejecta at the reverse shock front. Note that the radius of the circumstellar shock, $R_{\mathrm{cs}} $, is a factor 1.2-1.3 larger than $R_{\mathrm{s}}$ (\cite{CF94}).

The temperature behind the circumstellar shock is

\begin{equation}
\label{eq:Tcs}
T_{\mathrm{cs}}=2.27\times10^9\mu _{\mathrm{cs}}\frac{(\eta -3)^2}{(\eta -s)^2}V_{4}^{2}\ \ \ \mathrm{K},
\end{equation}

\noindent
where $V_{4}=V_{\mathrm{ej}}/10^4\ \mathrm{km\ s^{-1}}$. Values of $\mu $ and $\mu _{\mathrm{A}}$ are given in Table~\ref{tab:abund} for the different compositions considered in this paper. The composition of the circumstellar shock is the same as in the circumstellar medium, which we assume to be solar, with $\mu _{\mathrm{cs}}\approx 0.61$. The mass swept up by the outer shock front is

\begin{equation}
\label{eq:MswCS}
M_{\mathrm{cs}}=\frac{\dot{M}R_{\mathrm{s}}}{(3-s)v_{\mathrm{w}}}\left ( \frac{R_{\mathrm{s}}}{r_0}\right ) ^{(2-s)},
\end{equation}

\noindent
and the mass swept up by the reverse shock is

\begin{equation}
\label{eq:MswRev}
M_\mathrm{rev}=\frac{\eta -4}{4-s}M_{\mathrm{cs}}\ .
\end{equation}

The velocity of the reverse shock is

\begin{equation}
\label{eq:vRev}
V_{\mathrm{rev}}=\frac{3-s}{\eta -s}V_{\mathrm{ej}}\ ,
\end{equation}

\noindent
which in combination with the shock jump condition gives the velocity of the post-shock gas

\begin{equation}
\label{eq:vRev2}
v_0=\frac{V_{\mathrm{rev}}}{4}=\frac{(3-s)}{4(\eta -s)}V_{\mathrm{ej}}\ .
\end{equation}

As an alternative to using a self-similar model, dependent on the 
assumption that the ejecta density is a power law, $\rho \sim r^{-\eta }$, we have the option of using a
more detailed, non-power law model for the ejecta as 
input. In this case we use the thin shell approximation to solve for the position of the shock, and consequently the density, temperature and velocity of the
ejecta at this position (see e.g. Chevalier \& Fransson \cite{CF03}). The
post-shock values are then found from the shock jump conditions. Once the
state of the post-shock gas has been found, the treatment of
the interaction is the same in both cases, as described below.

An alternative scenario for the X-ray emission is shocks driven into dense circumstellar clumps (e.g. Chugai \cite{Chugai93}). In this case the X-ray emission would originate in radiative shocks within each clump. The structure of the radiative shocks would, however, for the same shock velocity be the same as for the reverse shock considered here. An especially interesting case of this type is the ring collision of SN~1987A (Michael et al.~\cite{Michael02}).

\subsection{Reverse shock structure}
The structure of the shocked ejecta behind the reverse shock is found
by solving the conservation equations. If we reduce the problem to one
dimension, which is justified for a thin shock, and assume steady state, these equations reduce to

\begin{equation}
\label{eq:hyd1}
\frac{\mathrm{d}}{\mathrm{d}x}\left(n_{\mathrm{i}}v\right)=0
\end{equation}

\begin{equation}
\label{eq:hyd2}
\frac{\mathrm{d}}{\mathrm{d}x}\left(\mu _{\mathrm{A}}m_{\mathrm{u}}n_{\mathrm{i}}v^2+p\right)=0
\end{equation}

\begin{equation}
\label{eq:hyd3}
\frac{\mathrm{d}}{\mathrm{d}x}\left[\left(\frac{1}{2}\mu _{\mathrm{A}}m_{\mathrm{u}}n_{\mathrm{i}}v^2+\frac{p}{\gamma -1}+p\right)v\right]=\Lambda(T) n_{\mathrm{i}}n_{\mathrm{e}} ,
\end{equation}

\noindent
where $n_{\mathrm{i}}$ is the ion number density and
$\Lambda $ is the cooling rate.  These
equations may be solved to give the structure of the radiative 
shell, once the values immediately behind
the shock are known. For the thin shell model the initial density and
temperature are given by Eqs.~(\ref{eq:densRev}) and (\ref{eq:Trev}),
and the initial velocity by Eq.~(\ref{eq:vRev2}). The initial pressure is
$p_{0}=(n_\mathrm{e,0}+n_\mathrm{0})kT_0$

The model is specified by the density, temperature and composition behind the shock. The initial value of the density depends on
which structure we assume for the ejecta (see Sect.~\ref{sect:intReg}). 
Once the density structure is set, the pressure and velocity of
each zone can be found from the continuity equations

\begin{equation} v=n_0v_{0}n_{\mathrm{i}}^{-1}
\end{equation}

\begin{equation} p=\mu _{\mathrm{A}}\mathrm{m_{u}}n_0v_{0}^{2}\left(1-\frac{n_0}{n_{\mathrm{i}}}\right)+p_{0},
\end{equation}

\noindent
and the temperature is given by 

\begin{eqnarray}
 T&=&{p \over k (n_{e}+n_{\mathrm{i}})} \nonumber \\
  &=&\frac{1}{k(n_{e}+n_{\mathrm{i}})}\left[(\mu _{\mathrm{A}} \mathrm{m_{u}}n_0v_{0}^{2}+p_{0})-\mu _{\mathrm{A}} \mathrm{m_{u}}\left(\frac{n_0v_{0}}{n_{\mathrm{i}}}\right)^2\right].
 \end{eqnarray}

The position of each zone is found from Eq.~(\ref{eq:hyd3}), using the density as the independent variable to solve for $x(n_{\mathrm{i}})$, 

\begin{equation}
\label{eq:Pos}
x =\int _{n_0}^{n_{\mathrm{i}}}  \frac{1}{\Lambda [T(n'),n']}  \left [\frac{2\mathrm{A}}{n'^5}+\frac{\mathrm{B}}{n'^4}\right ]\ \mathrm{d}n',
\end{equation}

\noindent
where $\Lambda $ is the total cooling rate, and A and B are constants, determined by the initial values of density and velocity behind the reverse shock

\begin{equation}
 \mathrm{A}=-\frac{(\gamma+1)}{2(\gamma -1)}\left(\frac{\mu _{\mathrm{A}}}{\mu }-1\right)\mu _{\mathrm{A}} \mathrm{m_{\mathrm{u}}}(n_0v_{0})^3,
 \end{equation}

\begin{equation}
 \mathrm{B}=\frac{\gamma }{(\gamma -1)}\left(\frac{\mu _{\mathrm{A}}}{\mu }-1\right)\left(\mu _{\mathrm{
A}}m_{\mathrm{u}}v_{0}^{3}+v_{0}T_0 \right) .
\end{equation}

\noindent
We continue the calculations until the temperature falls below $\sim 10^5$~K. Below this temperature the contribution to the X-ray emission is negligible. 

As the gas cools, the differential pressure between adjacent cooling regions causes instabilities, which can create secondary shocks within the cooling gas, particularly for high shock velocities (Chevalier \& Imamura, \cite{ChevIm82}; Plewa \& $\mathrm{R\acute{o}\dot{z}yczka}$~\cite{Plewa92}; Plewa~\cite{Plewa93}, \cite{Plewa95}; Dopita \& Sutherland, \cite{DopSuth96}; Sutherland et al., \cite{Suth03}). A complete treatment of this requires multidimensional hydrodynamical models. The specific case discussed in this paper was studied by Chevalier \& Blondin~(\cite{ChevBlond95}). Although the hydrodynamic structure can be severely affected, it is, however, likely that the time-averaged spectrum is close to that of a steady flow, as was indeed found by Innes~(\cite{Innes92}). As discussed by Sutherland \& Dopita (\cite{DopSuth96}), secondary shocks can nevertheless affect the line strengths of the lowest ionization species. This effect is of minor importance to us, since the lowest ionization stages have little or no impact on the X-ray emission. Dopita \& Sutherland (\cite{DopSuth96}) also conclude that limited spatial resolution in observational data has a more serious effect than neglecting thermal instabilities. Because several cooling cells and ranges of shock parameters are included in one beam, the effect of instabilities is reduced. This should certainly hold for the cases of interest here, where the size of each cooling cell is relatively small. Therefore, we conclude that neglecting instabilities in the cooling region  is not likely to have serious implications for the computed spectrum. The main effect should be that there should be some spread in the reverse shock velocity. The spectrum may then be a superposition of several radiative shocks with different velocities. This will also be the case for a reverse shock receding into a clumpy ejecta, as seen in e.g. Cas A (Chevalier \& Kirshner~\cite{ChevK79}; Chevalier \& Oishi~\cite{ChevOi03}; Fesen et al.~\cite{Fesen05}) and inferred for some supernovae, like SN~1995N~(Fransson et al.~\cite{Fransson2002}).

\section{Chemical composition}
\label{sect:composition}
The circumstellar shock sweeps up material lost by the star before the
explosion, which should be close to solar composition 
for Type II supernovae, although CNO burning may modify the relative abundances of C, N and O for massive progenitors (e.g., Fransson et al. \cite{Fransson2005}). The reverse shock, on the other hand,
propagates into the ejecta, sweeping up material, which may have undergone nucleosynthesis.  The composition of the ejecta varies
with radius, because of the successive
burning stages in the progenitor, although hydrodynamical mixing may
have modified this stratification. The mixing makes it more likely that enriched material will be
encountered at higher velocities. As the reverse shock moves backward
into the ejecta, it passes through regions of different composition. If the
progenitor has already lost part or all of its hydrogen envelope, this
may occur already during the early phases of the supernova. Examples
are the Type~IIb SN~1993J and some Type~IIn supernovae, like SN~1995N.

To investigate the effects of composition on the spectrum, we have calculated four models with the same temperature and maximum ejecta velocity, but different composition characteristics of the burning zones. One is for a solar composition, while the other three are characteristic for helium, carbon and oxygen dominated regions, respectively. Abundances of the latter three models were taken from the 4H47 model by Nomoto \& Hashimoto~(\cite{NomHash88}) and Shigeyama et al.~(\cite{Shigeyama94}) and are listed in Table~\ref{tab:abund}. The Ni abundances of the non-solar compositions, not given by the models, were estimated so that the Fe/Ni ratio would be the same as for the solar composition.

\begin{table*}[h*]
\caption{Fractional abundances by number and atomic weights for the different compositions.}
\label{tab:abund}
\begin{center}
\begin{tabular}{lcccc} \hline \hline 
Element & Solar & Helium zone & Carbon zone & Oxygen zone \ \\ \hline \\
H & $9.10\times 10^{-1}$& -- & --\ \\ 
He &$ 8.89\times 10^{-2}$& $9.82\times 10^{-1}$&$2.49\times 10^{-1}$&$2.36\times 10^{-10}$ \\ 
C &$3.79\times 10^{-4}$& $1.24\times 10^{-2}$&$4.29\times 10^{-1}$&$4.53\times 10^{-3}$\\ 
N & $7.92\times 10^{-5}$& -- & --&-- \\ 
O & $6.29\times 10^{-4}$&$3.75\times 10^{-3} $&$3.10\times 10^{-1}$&$8.49\times 10^{-1}$\\ 
Ne & $8.89\times 10^{-5}$&$1.56\times 10^{-3} $&$1.04\times 10^{-2}$&$2.32\times 10^{-4}$\\ 
Mg &$ 3.62\times 10^{-5}$&$ 8.24\times 10^{-5}$&$6.80\times 10^{-4}$&$9.29\times 10^{-2}$\\ 
Si &$3.46\times 10^{-5}$ & $1.02\times 10^{-4}$&$2.77\times 10^{-4}$&$3.59\times 10^{-2}$\\ 
S & $1.69\times 10^{-5}$& $4.91\times 10^{-5}$&$1.33\times 10^{-4}$&$1.58\times 10^{-2}$\\ 
Ar & $3.62\times 10^{-6}$& $1.92\times 10^{-5}$&$5.23\times 10^{-5}$& $1.11\times 10^{-3}$\\ 
Ca &$2.13\times 10^{-6} $& $6.08\times 10^{-6}$&$1.65\times 10^{-5}$&$2.54\times 10^{-5}$\\ 
Fe & $3.08\times 10^{-5}$&$1.22\times 10^{-4}$ &$3.33\times 10^{-4}$&$5.12\times 10^{-4}$\\
Ni & $ 1.62\times 10^{-6}$&$6.43\times 10^{-6}$&$1.75\times 10^{-5}$&$2.68\times 10^{-5}$\\ 
$\mu _{\mathrm{A}}$& 1.29 &  4.18 & 11.37 & 17.46 \\
$\mu $ & 0.61& 1.35 & \phantom{ } 1.70 &  \phantom{ } 1.80  \\
\hline
\end{tabular}
\end{center}
\end{table*}

\section{Ionization structure}
\label{sect:ionStruct}
In a medium with negligible external radiation the main contribution to
the ionization and recombination comes from collisions with thermal
electrons. Defining the ionization fraction of ionization stage $m$ of
element $Z$ as $X_{m}=n(Z_{m})/n(Z)$, we can write the rate of
change of $X_{m}$ in a Lagrangian reference frame as 

\begin{eqnarray}
\label{eq:ionbal}
\frac{\mathrm{d}X_{m}}{\mathrm{d}t} & = &n_{\mathrm{e}}[-(\alpha _m+C_m)X_{m} \nonumber \\
& & +\alpha _{m+1}X_{m+1}+C_{m-1}X_{m-1}] \ ,
\end{eqnarray}

\noindent
where $C_{m}$ is the collisional
ionization rate from ionization stage $m$ to $m+1$, while $\alpha _m$
is the recombination coefficient from ionization stage $m$ to $m-1$. Charge transfer, which is only important for temperatures below $\sim 10^5$~K, is ignored for the models in this paper. Charge transfer has, however, been included in a later version of the model, but was found to have no effect on the spectrum for the cases discussed in this paper.
The initial state of the gas behind the shock front depends on the state of ionization in the pre-shock gas, and is discussed in more detail below. We follow each gas element as it cools and recombines
after being shocked. In general, the recombination time will be longer
than the cooling time, leading to non-equilibrium ionization, with
highly ionized species being present at lower temperatures than in the
steady state case. This is taken into account by recognizing that each
position behind the shock corresponds to a specific time since the gas
entered the shock. The time that has passed since a particular zone, now at position $x$, was shocked, is then evaluated from $t=\int _{0}^{x}v(x')^{-1}dx'$.

Eq.~(\ref{eq:ionbal}) is solved for the 13 most abundant elements: H, He, C, N, O, Ne,
Mg, Si, S, Ar, Ca, Fe and Ni. References for the collisional ionization and recombination rates are given in the Appendix. 

In our models we neglect all external radiation, assuming that only shock heating is important for the ionization and emission. Since the radiation from the unshocked ejecta has a very low temperature compared to that of the shocked gas, this assumption is justified. Because we are only interested in the X-ray emitting gas, the effects of the X-rays on the cool shell can also be neglected. For the optical and UV radiation the reprocessing of the X-rays is, however, important~(\cite{CF94}).
In addition, the temperature is low enough, while
the density is high enough, for ion and electron temperatures behind
the shock to be nearly in equilibrium (\cite{FLC96}). We also neglect thermal conduction between the zones. This is discussed further in Sect.~\ref{sect:cond}.

The state of ionization immediately behind the shock is set by that of
the pre-shock gas. This is in turn determined by the X-ray photoionization from the shock. A self-consistent solution of
the two ionization states is therefore needed. As usual, the
photoionized gas is characterized by the photoionization parameter
$\zeta = L/n_e r^2$ (e.g., Tarter et al.~\cite{Tarter69}). Using Eqs.~(\ref{eq:densRev})
and~(\ref{eq:lum}) we find that

\begin{equation}
\zeta = 10.4\left( \frac{\mu \mu _{\mathrm{A}}}{\mu _{\mathrm{A}}-\mu}\right) \left({V_{\mathrm{rev}}\over 1000 \kms}\right)^3
\end{equation}

For a solar composition $\zeta =12.0(V_{\mathrm{rev}}/1000 \kms)^3$ and
for an oxygen dominated composition $\zeta = 20.9(V_{\mathrm{rev}}/
1000 \kms)^3$. The most important point to note here is that $\zeta$
only depends on the reverse shock velocity and the chemical
composition, but not on e.g., the mass loss rate parameter, $A_*$. The
pre-ionization does therefore not introduce any more parameters into
the problem. With $V_{\mathrm{rev}} \sim 1000 \kms$, implying $kT \sim 1$ keV,
and using the results of Tarter et al. (\cite{Tarter69}) for the ionization
structure based on a free-free spectrum, we expect that carbon will
mainly be in the form \ion{C}{v-vii}, oxygen mainly \ion{O}{vi-viii}, and neon
mainly \ion{Ne}{viii-ix}. This is confirmed by the models in Fransson~(\cite{Fransson1984}).

To calculate the pre-shock state we use the shock spectrum as input to
an updated version of the photoionization code in CF94. The result of
this is then used as input to the shock calculation. If necessary, the
new shock spectrum resulting from this can be used for a new
calculation of the pre-shock structure, etc. In practice, this
converges after the first iteration.

\section{Emission}
\label{sect:emission}

\subsection{Continuum emission}
We include free-free and free-bound emission, as
well as two-photon decay from H-like and He-like ions. The
emissivity (integrated over $4\pi$) for the free-free and free-bound continuum is given by (Sutherland \& Dopita,~\cite{SuthDop93})

\begin{eqnarray}
\label{eq:emissivity}
\epsilon_{\lambda ff,fb} &=&2.05 \times 10^{-19}n_{\mathrm{e}}^{2}G_{\mathrm{c}}\lambda^{-2} \nonumber \\
& &\times \mathrm{e}^{-E/kT} T^{-1/2}\ \ \ \ \ \ \mathrm{ergs\ s^{-1}cm^{-3}\AA ^{-1}},   
\end{eqnarray}

\noindent
where the total Gaunt factor $G_c$ is the sum of the Gaunt factors for
the free-free and free-bound transitions, and is written as

\begin{equation}
G_{\mathrm{c}}=G_{\mathrm{ff}}+G_{\mathrm{fb}}.
\end{equation}

Sutherland \& Dopita~(\cite{SuthDop93}) and Mewe et al.~(\cite{Mewe86}) include a Gaunt factor for the two-photon emission as well. For this paper we evaluate the two-photon contribution explicitly from the level populations, as described below. Any quenching of this at high densities is therefore taken into account. The bound-free Gaunt factor is evaluated according to the
method of Mewe et al. (\cite{Mewe86}), which we follow also for the free-free
emission,

\begin{equation}
G_{\mathrm{ff}}=\frac{n_\mathrm{H}}{n_{\mathrm{e}}}\sum _{Z,i}A_ZX_{m}(Z)z_{n_{0}}^{2}g_{\mathrm{ff}}, 
\end{equation}

\noindent
where $A_Z$ is the abundance of element $Z$ and $z_{n_{0}}$ is the effective charge of the ground state, defined by 

\begin{equation}
  E_{\mathrm{ion}}^{m-1}(Z)=E_{\mathrm{ion}}(\mathrm{H})\frac{z_n^2}{n^2}.
  \end{equation}

\noindent
Here, $n$ is the principal quantum number, with the modification that we use the calculations of
Sutherland (\cite{Suth98}) for the individual free-free Gaunt factors, $g_{\mathrm{ff}}$. 

For the two-photon emission, we calculate the emissivity as 

\begin{equation}
\epsilon_{\lambda 2\gamma }=\sum _{Z,z}n_{2\gamma }(Z,z)A_{2\gamma }(Z,z)E_{0}\phi (y),
\end{equation}

\noindent
where the sum goes over H-like and He-like ions, $z$, of all elements $Z$. Further, $n_{2\gamma }$ is the density of the upper level of two-photon transitions in ion ($Z,z$) (either $2s^2S_{1/2}$ or $2s^1S_{0}$), $E_{0}$ is the energy difference between the levels and $\phi (y)$ is the normalized distribution function of the total two-photon emission which comes out at energy $y$, with $y=E/E_{0}$. For H-like ions we use the distribution function of Goldman \& Drake~(\cite{GoldDrake81}), while we use data from Drake et al.~(\cite{Drake69}) for He-like ions.
The total emissivity is then

\begin{equation}
 \epsilon_{\lambda }=  \epsilon_{\lambda ff,fb} + \epsilon_{\lambda 2\gamma }.
 \end{equation}

\subsection{Line emission}
In a collisionally ionized, low density plasma the main processes responsible for
populating the excited levels are collisional and radiative excitation and de-excitation. 
We can then write the rate of
change of the population $n_{m,k}$ of level $k$ of ion $m$ as

\begin{eqnarray} 
\label{eq:levelPop}
\frac{\mathrm{d}n_{m,k}}{\mathrm{d}t} & = & -n_{m,k}n_{\mathrm{e}}(\alpha _{m,k}+C_{m})\nonumber \\
& - & n_{m,k}\sum_{j<k}(A_{kj}\beta _{kj}+C_{kj}n_{\mathrm{e}}) \nonumber \\
& - & n_{m,k}\sum_{j>k}C_{kj}n_{\mathrm{e}}+\sum_{j<k}n_{m,j}C_{jk}n_{\mathrm{e}} \nonumber \\
& + & \sum_{j>k}n_{m,j}(A_{jk}\beta _{jk}+C_{jk}n_{\mathrm{e}}) \\ \nonumber
& + & n_{m+1}\alpha _{m+1,k}n_{\mathrm{e}}+n_{m-1}n_{\mathrm{e}}C_{m-1}\ ,
\end{eqnarray}

\noindent
where $A_{jk}$ is the transition probability, $\beta
_{kj}$ is the escape probability, $\alpha_{m+1,k}$ is the
recombination rate from ion $m+1$ to ion $m$, level $k$ and  $C_{m-1}$ is the collisional ionization rate from ion $m-1$ to ion $m$. $C_{kj}$ and $C_{jk}$ are the rate coefficients of collisional excitation and de-excitation, respectively, with

\begin{equation}
 C_{kj}=\frac{g_{j}}{g_{k}}C_{jk}\mathrm{e}^{-(E_{j}-E_{k})/kT}  \ \ \ \ \ \ j>k,
\end{equation}

\noindent
where $g_{k}$ and $g_{j}$ are the statistical weights of levels $k$ and $j$, respectively.

Radiative recombination to individual levels is included for He-like, Li-like and Na-like ions, and dielectronic recombination to \ion{Fe}{xvii} and \ion{Fe}{xviii}. For the ions where radiative recombination to individual levels is included we scale the rates to the individual levels with the total recombination to the ion, which includes the dielectronic recombination, thus distributing this contribution among the levels. We find that recombination to individual levels, where it is included, has little effect on the level populations.
We assume all collisional ionization to take place to and from the ground states. Similarly, recombination from ion $m$ to $m-1$ is assumed to only occur from the ground state, so that $\alpha _{m,k}=0$ for $k>0$.

In the Sobolev approximation the escape probability is calculated from

\begin{equation}
\label{eq:escProb}
\beta _{kj}=\frac {1-\mathrm{e}^{-\tau _{kj}}}{\tau _{kj}},
\end{equation}

\noindent
where the optical depth of the transition $k\rightarrow j$, $\tau _{kj}$, is given by

\begin{equation}
\label{eq:optDepth}
\tau _{kj}=n_{m}\sigma _{kj}\mathcal{L}_{\mathrm{s}}\left(1-\frac{g_kn_{m,j}}{g_jn_{m,k}}\right).
\end{equation}

\noindent
Here $n_{\mathrm{m}}$ is the density of ion $m$, $\sigma _{kj}$ the cross section for the transition, given by

\begin{equation}
\label{eq:crossSect}
\sigma _{kj}=\frac{A_{jk}n_{m,k}\lambda _{0}^{3}}{8\pi v_{\mathrm{th}}}\frac{g_j}{g_k},
\end{equation}

\noindent
and $\mathcal{L}_{\mathrm{s}}$ is the Sobolev length, defined as

\begin{equation}
\mathcal{L}_{\mathrm{s}}=v_{\mathrm{th}}\frac{\mathrm{d}r}{\mathrm{d}v},
\end{equation}

\noindent
where $v_{\mathrm{th}}$ is the thermal velocity. The Sobolev approximation holds as long as the size of the emitting region is larger than the Sobolev length, $\mathcal{L}_{\mathrm{s}}$. In the opposite case we use the physical thickness of the region instead of $\mathcal{L}_{\mathrm{s}}$. As an estimate of $dv/dr$ we take the velocity gradient from the shock to where the temperature has fallen to half of its initial value. This approximation does not hold for low temperatures, when the velocity gradient steepens. However, the net effect is that we overestimate the absorbing length, and thus the optical depth, and since we find that even in this case optical depth effects are not important (see Sect.~\ref{sect:optDepth}), the errors introduced by this method are negligible. When computing the Sobolev length for lines formed in a given shell, we use the thermal velocity in that shell.

Once the level populations have been found, the volume emissivity, $\epsilon _{kj}$, in
each line is computed according to

\begin{equation}
\label{eq:lineEm}
\epsilon_{kj}=n_{m,k}A_{kj}E_{kj}\beta _{kj},
\end{equation}

\noindent
where $E_{kj}$ is the energy difference between levels $k$ and $j$.

A list of all ions for which level populations are calculated is given in Table~\ref{tab:ionLev} in the Appendix, as well as references for the atomic data. The majority of the atomic data have been taken from version 4.2 of the Chianti database~(Dere et
al.~\cite{Dere97}; Young et al.~\cite{Young03}). After this paper had been submitted, we tested the code with version 5.1 of Chianti~(Landi et al.~\cite{Landi05}), but it gave only negligible effects. To speed up the calculations, only the level populations for ions which have an appreciable abundance are calculated in each zone.

The total cooling per unit volume in Eq.~(\ref{eq:Pos}) is calculated by adding the contributions from line emission and continuum processes

\begin{equation}
\label{eq:cooling}
\sum _{\lambda}\epsilon_{\lambda}+\sum _{k,j}\epsilon_{kj}=n_{\mathrm{e}}n_{\mathrm{i}}\Lambda (T_{\mathrm{e}}) .
\end{equation}

\subsection{Conduction}
\label{sect:cond}
As mentioned previously, we have neglected thermal conduction in the cooling region. Previous studies have indicated that while this is a reasonable approximation for cosmic abundances, a higher metallicity will increase the importance of conduction (Borkowski et al.~\cite{Bork89}; Borkowski \& Shull~\cite{Bork90}). 

The importance of conduction may be estimated from the ratio between
the conduction length scale and the cooling length scale, $\alpha
=L_{\mathrm{cond}}/L_{\mathrm{cool}}$ (Borkowski et
al.~\cite{Bork89}). The conduction length gives an estimate of how far heat can be conducted, while the cooling length determines the slope of the temperature gradient. Consequently, conduction dominates if the conduction length exceeds the cooling length, i.e. if $\alpha \ga 1$. We estimate this parameter directly behind the
shock in the solar and  the oxygen dominated models. For solar composition we find $\alpha \ll 1$, while for the
oxygen model $\alpha \sim 0.02$, which justifies neglect of conduction
even in this case. 

This conclusion is in agreement with Borkowski et
al. (\cite{Bork89}) for the solar model, but not for the oxygen
model. They argue that for a pure oxygen model the cooling rate, and therefore also $\alpha $, is increased by a factor $\sim 1000$ due partly to the higher metallicity, but mainly due to non-equilibrium cooling. The latter is, however, an effect of the low shock velocities in their models, which means that the pre-shock gas is not fully ionized. Therefore, excitation and ionization of neutrals strongly enhance the post-shock cooling, thereby reducing the cooling length. This is not the case in our models, where the shock velocity is high enough for the pre-shock gas to be fully ionized. Instead, non-equilibrium cooling tends to lower the cooling rate compared to the equilibrium case, since high ionization stages are present at lower temperatures where they cool less efficiently. Thereby the non-equilibrium cooling lessens the importance of conduction in our models.
We therefore feel justified in neglecting thermal conduction
in the models discussed in this work, but keeping in mind that for
high metallicities, conduction may become important for certain shock
parameters, as it would be e.g., in supernova remnants where Fe-rich
material is encountered (e.g., Sorokina et al. \cite{Soro04}). It is, however, unlikely to change our qualitative conclusions. In addition, any magnetic field present is likely to decrease the conductivity.

\subsection{Line profiles}
\label{sect:lineProf}
The main process responsible for broadening of the emission lines is the
expansion of the interaction shell, which has a velocity of $V_{\mathrm{s}}\sim
10^4\ \mathrm{km\ s^{-1}}$. In the thin shell approximation we may assume $V_{\mathrm{s}}=V_{\mathrm{ej}}$. For a thin shell at radius $r_s$, the line profile of an optically thin line is flat out to the velocity of the shell. For lines which are optically thick,
the line profile depends on the velocity gradient. For $V\propto r$
the line is flat, while for constant velocity it is parabolic
(e.g., Ambartsumian \cite{Ambartsumian57}). A breakdown of the Sobolev
approximation, as discussed above, may lead to double peaked line
profiles for the optically thick lines, as seen for the H$\alpha $
line in SN~1998S ({Fransson et al. \cite{Fransson2005}). 

In addition, the ejecta will at least during the first few years be optically thick to X-rays from the reverse shock. Therefore, only the emission from the approaching part of the shock will in general be observed. 

Because the line profile in X-rays will usually be dominated by the instrumental profile, we here use, for simplicity, a flat, box-like profile in all cases.

\section{Results}
\label{sect:discussion}
\subsection{Model parameters}
\label{sect:param}
The model described here is applicable to supernovae where the reverse
shock is radiative. Progenitors of Type II supernovae have mass loss rates in the
range $10^{-6} - 10^{-4}\ \mathrm{M_{\odot}\ yr^{-1}}$. For a wind
velocity $v_{\mathrm{w}}=10\ \mathrm{km\ s^{-1}}$, this gives
$\tilde{A}_{*}\approx 0.1-10$, with Type~IIPs in the lower range, and Type~IILs and Type~IIns having $\tilde{A}_{*}\ga 1$. As we will discuss in Sect.~\ref{sect:timescales}, for $\eta \ga 10$ the reverse
shock is then radiative to at least a hundred days for solar
composition and a shock velocity of $\la 10^4\ \mathrm{km\ s^{-1}}$. For
steeper density gradients, and for helium or oxygen rich ejecta, the
radiative phase will be longer. The steady state condition for
the reverse shock is, therefore, at least for the first few hundred days, fulfilled
for most Type~IIL and Type~IIn supernovae. 

Type~IIP supernovae are,  because of the moderate mass loss rates, in general adiabatic, and are discussed in Chevalier et al.~(\cite{CFN05}). Type Ib/c supernovae are thought to originate from Wolf-Rayet stars, which have
much faster stellar winds, with 
$\tilde{A}_{*}\sim 10^{-3}$. Therefore, the shock is not
radiative beyond a few days or even hours, and consequently our models
are not applicable for Type Ib/c supernovae, unless the density
gradient is very steep. For these supernovae a better
assumption is to employ an adiabatic model for the temperature and
density. Inverse Compton scattering of
photospheric photons by relativistic electrons may also contribute to the X-ray emission from Type Ib/c supernovae 
(Bj\"ornsson \& Fransson 2003). Also inverse Compton scattering by thermal
electrons behind the circumstellar shock may under certain conditions
contribute (Sect. \ref{sect:csshock}).

The parameters which characterize a given model are the temperature,
composition and density immediately behind the reverse shock. Further,
for a stationary shock the density is mainly important for the
absolute flux, not for the spectral shape, as long as optical depth
effects and collisional de-excitation are not important, which is in
most cases a good approximation (Sect.~\ref{sect:optDepth}). As we discuss below,
departures from ionization equilibrium are only important at low
temperatures, and do not influence the spectrum above 100~eV. Altogether, this implies that \emph{the
shock spectrum is only sensitive to the reverse shock
velocity and the composition. }

For the models discussed in this paper, we take
$\tilde{A}_{*}=1$, $V_{4} =1$ and $s=2$, and a time of 100 days after
explosion, unless otherwise noted. In order to study the effects of different reverse shock temperatures or, alternatively, different chemical compositions, the density gradient of the ejecta is varied according to Eq.~(\ref{eq:Trev}), as we vary either the temperature or the composition. We stress, however, that although the density behind the reverse shock varies with $\eta $, these parameters are only important for the scaling of the absolute flux. 

The normalization of the spectrum is given by the fact that the total luminosity from a radiative shock is

\begin{eqnarray}
\label{eq:lum}
L_{\mathrm{rev}}&=&4\pi R_{\mathrm{s}}^{2}\frac{1}{2}\rho
_{\mathrm{ej}}V_{\mathrm{rev}}^{3}   \nonumber \\
&=&1.6\times
10^{41}\frac{(\eta-3)(\eta-4)}{(\eta-2)^3}\tilde{A}_{*}V_{4}^{3} \ \ 
\mathrm{erg\ s^{-1}},
\end{eqnarray}

\noindent
if $s=2$.  Because the velocity depends on time as
$V\propto t^{-1/(\eta -2)}$, this means that the total luminosity of
the reverse shock decreases slowly with time, as long as $\eta $ is
constant. However, for the \emph{observed} flux this is more than compensated
by the decreasing optical depth in the cool, absorbing shell, which lets
through more of the emitted radiation as the region expands (see
Sect.~\ref{ssect:absorption}). Note also that approximately half of the above luminosity is absorbed by the ejecta (Sect.~\ref{sect:lineProf}, \ref{sect:csshock}).

\subsection{Shock structure}
\label{ssect:shockStruct}
Radiative shocks occur in a wide range of circumstances in the
interstellar medium, and have been discussed by a
number of authors. The first numerical models of cooling shocks were
made by Cox~(\cite{Cox72}), who considered a planar shock with a
velocity of $100\ \mathrm{km\ s^{-1}}$ and applied the results to the
Cygnus Loop. Later, improved models were developed by
Raymond~(\cite{Ray79}), Dopita~(\cite{Dopita76},
\cite{Dopita77}) and others (see also Dopita \& Sutherland~\cite{DopSuth96} and
references therein). These models were intended for such diverse
phenomena as old supernova remnants and narrow line emission regions
in active galactic nuclei, but are in general not applicable to young
supernovae. The reason for this is that for the low densities in
supernova remnants the shock is not radiative above $\sim 200\
\mathrm{km\ s^{-1}}$.  

The effects of composition on the properties of the cooling gas in radiative shocks have been studied by Itoh (\cite{Itoh81a}, \cite{Itoh81b}, \cite{Itoh86}, \cite{Itoh88}), Borkowski et al.~(\cite{Bork89}) and Borkowski \& Shull~(\cite{Bork90}). The main difference between these studies and the present one is again the lower velocity, $\la 200\ \mathrm{km \ s^{-1}}$, and density in their models, which is the reason for the somewhat differing results, e.g. as regards the role of conduction, as mentioned earlier. 

The hydrodynamic evolution of fast shocks evolving in a dense medium, applicable to AGNs and young supernova remnants, was calculated by Plewa \& $\mathrm{R\acute{o}\dot{z}yczka}$~(\cite{Plewa92}) and Plewa~(\cite{Plewa93}, \cite{Plewa95}). While these models could indeed be applicable to supernovae if the composition is close to solar, their treatment of the cooling in a time-independent formalism makes it less applicable to the reverse shock. As  we discuss below, a higher metallicity, as appropriate for the reverse shock, leads to a stronger cooling, and thereby a need for a time-dependent calculation of the ionization balance and cooling. They also treat the emission in a simplified manner, assuming collisional ionization balance and computing the emission in the cooling region from given line intensities. In contrast, we compute the emission from transitions in multi-level ions, and couple the resulting cooling to the hydrodynamics resulting in detailed and self-consistent spectra.

An important feature of high velocity shocks is
the creation of ionizing photons, which ionize the pre-shock gas and
the cooling, shocked gas (Sect.~\ref{sect:ionStruct}). If neutrals are present, some time is needed
for the shocked gas to become ionized. As the shocked gas approaches ionization equilibrium
corresponding to the post-shock temperature, the
non-equilibrium ionization state creates strong line emission, which
causes rapid cooling. For very fast
shocks ($V_{s}\ga 500~\mathrm{km\ s^{-1}}$), like the shocks considered
in the present paper, the gas is, however, nearly fully pre-ionized
(Sect.~\ref{sect:ionStruct}, Fransson~\cite{Fransson1984}; \cite{CF94}).

When ionization equilibrium has been reached, the gas cools slowly, with
the ionization state adjusting to equilibrium at the local
temperature more rapidly than the cooling occurs. This is where most of the emission occurs, because of the high temperature. When the local
temperature falls below that corresponding to the maximum of the
cooling function, around $10^6$~K, non-equilibrium ionization sets in. The gas cools rapidly, until the temperature falls below $10^4$~K, where photoionization heating balances cooling.  

The gas close to the contact discontinuity has a temperature $\la 10^{4}$~K and a density of $\sim (10^7~\mathrm{K}/10^4~\mathrm{K})\sim 10^3$ times that of the post-shock gas, i.e., $10^{10}-10^{11}\ \mathrm{cm}^{-3}$, and absorbs a large fraction of the  X-rays. (Sect.~\ref{ssect:absorption}). A large fraction of the optical and UV emission originates in the cold, dense  shell (\cite{CF94}).

For a given reverse shock velocity and composition, the post-shock temperature will be lower the steeper
the ejecta density gradient is, but is typically $\sim 10^7$~K. This is close to the minimum of the cooling function, so in models with a temperature lower than $\sim 10^7$~K a cooling instability
will set in almost immediately, whereas models with a higher
temperature will take longer before cooling starts, i.e., the
interaction region will be thicker. 

\begin{figure}[h]
\begin{center}
\resizebox{\hsize}{!}{\includegraphics{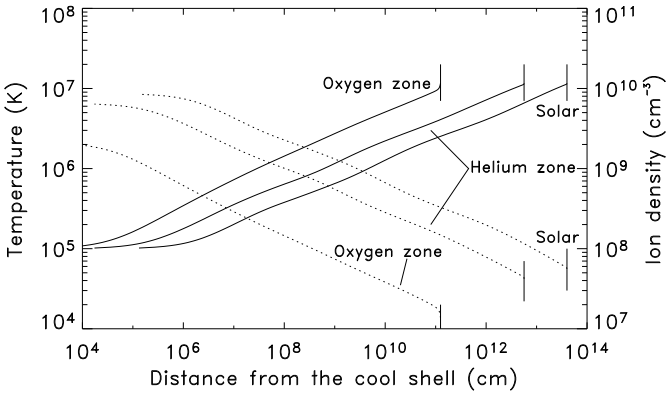}}
\end{center}
\caption{Density (dashed lines) and temperature (solid lines) behind the reverse shock as a function of distance from the contact discontinuity in a model with $V_{4} =1$, $s=2$, $\tilde{A}_{*}=1$, and $T_0=1.0$~keV at 100 days after the explosion and for three compositions. In these models the radius of the interaction region is $R_{\mathrm{s}}=8.64\times 10^{15}$ cm. The vertical lines indicate the position of the shock with respect to the cool shell in the three cases.} 
\label{fig:tempDens}
\end{figure}

In Fig.~\ref{fig:tempDens} we show the density and temperature structure for our standard model with $V_{4} =1$, $\tilde{A}_{*}=1$ and $T_0=1.0$~keV at 100 days after the explosion, and for three compositions. To show the drop in temperature and ionization close to the cool shell more clearly, we plot this \emph{as the distance from the end of the cooling instability}, rather than from the shock.

An analytical estimate of the thickness of the cooling region may be obtained from
the cooling time and the velocity of the post-shock gas, $V_{\mathrm{rev}}/4$, so that $D= t_{\mathrm{cool}}V_{\mathrm{rev}}/4$ where

\begin{equation}
\label{eq:tcool1}
t_{\mathrm{cool}}=\frac{3kT(n_{\mathrm{e}}+n_{\mathrm{i}})}{2\Lambda n_{\mathrm{e}}n_{\mathrm{i}}}=\frac{3kT}{2(1-\mu /\mu _{\mathrm{A}})n_{\mathrm{i}}\Lambda}\ .
\end{equation}

The equilibrium cooling function may be fit by a function of the form

\begin{equation}
\label{eq:coolFunc}
\Lambda =AT_{6}^{-\alpha} + BT_{6}^{\beta},
\end{equation}

\noindent
where $T_{6}=T/10^6\ \mathrm{K}$ and the fitting parameters A, B, $\alpha $ and $\beta $ depend on the composition of the gas, and are given in Table~\ref{tab:coolingPar} for the compositions considered here. 
In the range $10^{5}-10^{9}$~K, which are the temperatures contributing to the X-ray emission,
the fits are in general valid to within a factor of 2. The exceptions are the carbon and oxygen models, for which the cooling below $10^6$~K varies strongly. The fits to the cooling functions are plotted in Fig.~\ref{fig:coolCurve}, together with the cooling functions from our model calculations, assuming ionization equilibrium.

\begin{figure}[h]
\begin{center}
\resizebox{\hsize}{!}{\includegraphics{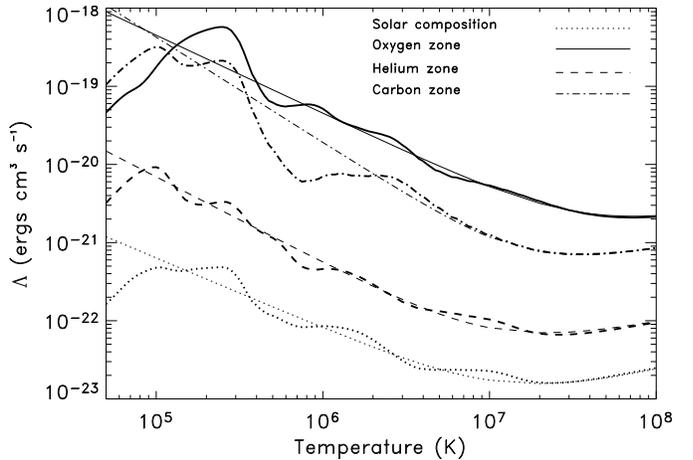}}
\end{center}
\caption{Cooling curves for four compositions. Cooling functions from the model calculations for each composition, assuming ionization equilibrium,  are plotted with the corresponding analytical fit, with the fitting parameters given in Table~\ref{tab:coolingPar}.} 
\label{fig:coolCurve}
\end{figure}

For the temperatures of interest, only the first term of the cooling function in Eq.~(\ref{eq:coolFunc}) contributes. Combining this with Eq.~(\ref{eq:tcool1}), and using Eqs.~(\ref{eq:Trev}) and (\ref{eq:Tcs}) for the temperature, we can express the thickness of the region as  

\begin{eqnarray}
\label{eq:deltaX}
D &\sim  & \frac{1.45\times 10^{-8}}{A}\frac{\mu _{\mathrm{A}}^{2}\mu ^{1+\alpha }}{(\mu _{\mathrm{A}}-\mu )}  \nonumber \\
& & \times \frac{\left(2.27\times 10^{3}\right)^{\alpha}}{(\eta -3)(\eta -4)(\eta -2)^{3+2\alpha }}\frac{t_{d}^{2}}{\tilde{A}_{*}}V_{4}^{5+2\alpha }  \nonumber \\
& = & \frac{D_{0}}{(\eta -3)(\eta -4)(\eta -2)^{3+2\alpha }}\frac{t_{d}^{2}}{\tilde{A}_{*}}V_{4}^{5+2\alpha } \ ,
\end{eqnarray}

\noindent
where the time is in days since the explosion, and we have set $s=2$.  The composition dependent constant $D_{0}$ is given in Table~\ref{tab:coolingPar} for the four compositions considered here.

For the model parameters above ($V_{4}=1$, $\tilde{A}_{*}=1$, $T_{0}=1$~keV),  the analytical estimate gives $D\sim 2.1\times 10^{14}$~cm, for solar composition, while the model calculations give $D\sim 5.3\times 10^{13}$~cm. The factor of four difference results from the approximate form of the cooling function in Eq.~(\ref{eq:coolFunc}) and the fact that the cooling increases as the temperature drops, leading to a collapse of the region. 
For the non-solar models the analytical estimates give $D\sim 3.5\times 10^{13}$~cm for the helium model, $D\sim 2.8\times 10^{12}$~cm for the carbon model and $D\sim 6.7\times 10^{11}$~cm for the oxygen model. The numerical results for these models are $D\sim 7.8\times 10^{12}$~cm, $D\sim 9.4\times 10^{11}$~cm and $D\sim 3.1\times 10^{11}$~cm, respectively. In all cases the thickness of the cooling gas is much smaller than the shock radius, justifying the thin shell and stationary approximations.

\begin{table*}[h*]
\caption{Fitting parameters for the cooling function of Eq.~(\ref{eq:coolFunc}) and the thickness parameter, $D_{0}$ of the cooling region from Eq.~(\ref{eq:deltaX}).}
\label{tab:coolingPar}
\begin{center}
\begin{tabular}{lccccc} \hline \hline 
Composition & A ($\mathrm{erg\ cm^3\ s^{-1}}$) & B ($\mathrm{erg\ cm^3\ s^{-1}}$) & $\alpha $ & $\beta $ & $D_{0} $ (cm) \ \\ \hline \\
Solar & $8.0\times 10^{-23}$& $2.3\times 10^{-24}$ & 0.90 & 0.50 & $1.82\times 10^{17}$\ \\ 
Helium zone &$ 5.5\times 10^{-22}$& $1.5\times 10^{-23}$& 1.10 & 0.40 & $1.50\times 10^{18}$ \\ 
Carbon zone &$1.9\times 10^{-20}$& $1.3\times 10^{-22}$&1.35&0.40&$1.20\times 10^{18}$\\ 
Oxygen zone & $4.5\times 10^{-20}$& $ 2.8\times 10^{-22}$ & 1.00 & 0.40 & $4.61\times 10^{16}$ \\ 
\hline
\end{tabular}
\end{center}
\end{table*}

The ionization structure
of Fe in the cooling region behind the reverse shock for the solar composition model is plotted in
Fig.~\ref{fig:iondensFe}. We see that Fe is never fully ionized, because of the comparatively low shock temperature. Immediately behind the shock front \ion{Fe}{xxi} is most abundant, with noticeable fractions of \ion{Fe}{xvii--xxv}. These ions dominate the bulk of the cooling gas, but further downstream also lower ionization stages become sufficiently abundant to affect the emission. 

\begin{figure}[h]
\begin{center}
\resizebox{\hsize}{!}{\includegraphics{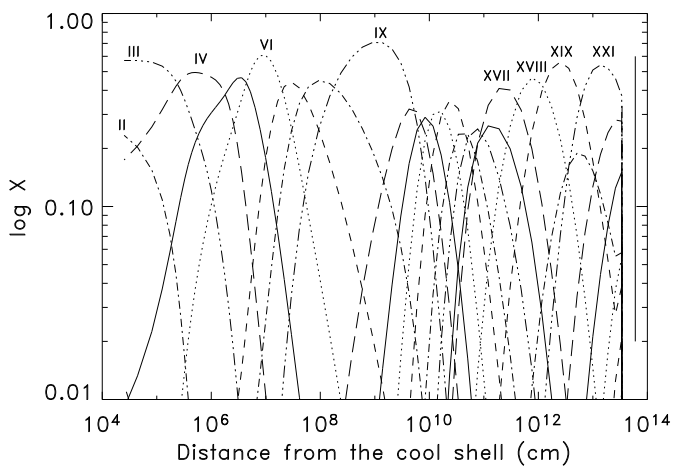}}
\end{center}
\caption{Ionization structure of Fe behind the reverse shock for the solar composition model in Fig.~\ref{fig:tempDens}. The vertical line indicates the position of the shock.} 
\label{fig:iondensFe}
\end{figure}

\subsection{Cooling and ionization timescales}
\label{sect:timescales}
The main assumption of our model is that the flow behind the reverse shock is stationary, i.e., that the time needed for a process to occur is short
compared with the expansion time. In this case we can calculate the temporal
evolution assuming steady state with given values of $n_0$ and $T_0$ for each epoch. For the hydrodynamic equations the relevant timescale is the cooling
timescale, given by Eq.~(\ref{eq:tcool1}). 

We check the assumption of a stationary flow by
estimating $t_{\mathrm{cool}}$ in all zones. Using Eq. (\ref{eq:densRev2}) for the ion density and s=2, we can express the ratio of the cooling and expansion timescales as

\begin{equation}
\frac{t_{\mathrm{cool}}}{t} =  \frac{6.7\times 10^{-22}\mu _{\mathrm{A}}^2\mu V_{4}^4 t_{\mathrm{d}}}{(\mu _{\mathrm{A}}-\mu )(\eta -2)^2(\eta -3)(\eta -4)\Lambda (T) \tilde{A}_{*}},
\label{eq:tcool}
\end{equation}

\noindent
where $t_{\mathrm{d}}$ is the time in days since the explosion. The cooling time is thus determined by the shock velocity, $V_{s}$, and the ratio $t_{\mathrm{d}}/\tilde{A}_{*}$. 

In Fig.~\ref{fig:shockVelTime} we plot this relation for $s=2$, three different values of $\eta $ and three of the chemical compositions in Table~\ref{tab:abund}. For a particular value of $\tilde{A}_{*}$ the reverse shock is radiative at a given time, $t_{\mathrm{d}}$, if the velocity of the outer shock is below the curve corresponding to the ejecta density profile. We see that the higher the value of $\eta $, the larger the range of shock velocities for which the reverse shock becomes radiative early on. The same is true as we go from solar composition to more metal rich gas.

\begin{figure}[h]
\begin{center}
\resizebox{\hsize}{!}{\includegraphics{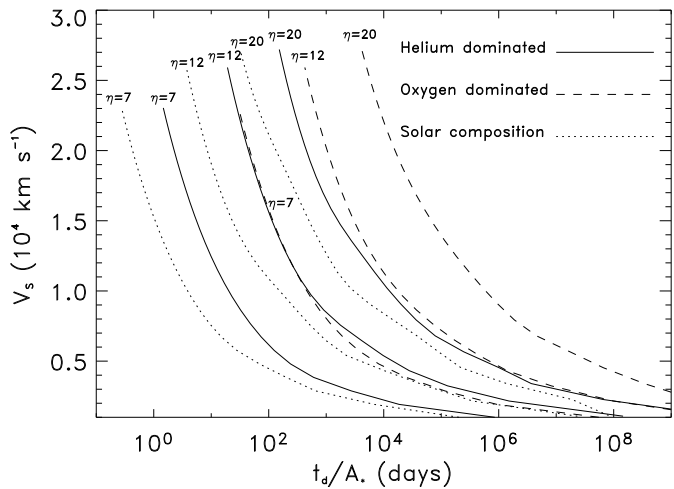}}
\end{center}
\caption{Ratio of cooling time ($t_{\mathrm{cool}}$) and age ($t_{\mathrm{d}}$) of the shocked ejecta. The lines represent the limiting velocities for which the
cooling time behind the reverse shock is equal to the expansion time. If the velocity of the
circumstellar shock at a given time is 
lower than values along the curve corresponding to the relevant density slope, $\eta $,
the reverse shock is radiative. If it is higher, the shock is
adiabatic. } 
\label{fig:shockVelTime}
\end{figure}

Due to the nearly constant pressure behind the reverse shock, $n_{\mathrm{e}}\propto T_{\mathrm{e}}^{-1}$ in the post-shock region. The density in a model with $T_0=10^7$~K will therefore increase by a factor $\sim 10^2$ before reaching $T_{\mathrm{e}}\sim 10^5$~K. Therefore, if the assumption of  stationarity holds immediately behind the reverse shock, it will be a good approximation in all of the post-shock gas.

Although it has been found that in a supernova remnant the ionization
structure must be modeled by a time-dependent formalism (e.g., Kaastra
\& Jansen~\cite{KaastraJansen93}; Sorokina et al. \cite{Soro04}), it is not clear if this is also the case for a young supernova, where the  densities are much higher. We will therefore next explore the differences between steady state and time dependent calculations for the ionization balance for a variety of parameters and compositions (Table~\ref{tab:abund}). 

For the ionization balance, the requirement is that  the collisional ionization parameter $\tilde{t}_{\mathrm{rel}}=n_{\mathrm{e}}t_{\mathrm{rel}}$ be sufficiently small compared to the expansion and cooling timescales. The relaxation time for a particular ion, $t_{\mathrm{rel}}$, is the shortest of the ionization and recombination times (e.g., Mewe~\cite{Mewe99}), given by
$t_{\mathrm{ion}}=[C(T_{\mathrm{e}})n_{\mathrm{e}}]^{-1}$ and $t_{\mathrm{rec}}=[\alpha(T_{\mathrm{e}})n_{\mathrm{e}}]^{-1}$, where $C(T_{\mathrm{e}})$ and $\alpha (T_{\mathrm{e}})$ are the collisional
ionization and recombination coefficients, respectively, for a given ion at a specific temperature. There is,  therefore, a specific, temperature dependent relaxation parameter for each ion, which can be expressed as $\tilde{t}_{\mathrm{rel}}=\mathrm{min}[1/C(T_{\mathrm{e}}),1/\alpha (T_{\mathrm{e}})]$. In Fig.~\ref{fig:ionTime} we show how this parameter for \ion{Fe}{ii--xxvi} varies with temperature for $T\ga 10^5$~K. At high temperatures it is given by the ionization timescale, while at lower temperatures it is determined by the recombination time. The relaxation parameter depends only on atomic physics. 

Let us first consider the ratio of $\tilde{t}_{\mathrm{rel}}$ to the expansion timescale, $t$. For a given temperature, a limiting density above which steady state applies may be found from

\begin{equation}
\label{eq:densReq}
n_{\mathrm{e}}t\ga \tilde{t}_{\mathrm{rel}}(T) \ .
\end{equation}

\noindent
Using the ion density from Eq.~(\ref{eq:densRev2}), writing the electron density as $n_{\mathrm{e}}=n_{\mathrm{ion}}(\mu_{\mathrm{A}}-\mu )/\mu $, and assuming s=2, this relation translates into the requirement that the ejecta density gradient satisfy the following relation

\begin{eqnarray}
\label{eq:densGradReq}
(\eta -3)(\eta -4) & \ga &0.12\frac{\mu \mu _{\mathrm{A}}}{\mu _{\mathrm{A}}-\mu }
\frac{V_{4}^{2}}{\tilde{A}_{*}} \nonumber \\
& &\times \left(\frac{\tilde{t}_{\mathrm{rel}}}{10^{7}\ \mathrm{days\ cm^{-3}}}\right)\left(\frac{t_{\mathrm{d}}}{100 \ \mathrm{days}}\right).
\end{eqnarray}

\noindent
The highest values of the relaxation parameter in Fig.~\ref{fig:ionTime} are those of \ion{Fe}{xxv} and \ion{Fe}{xxvi} at $T=10^8$~K, where $\tilde{t}_{\mathrm{rel}}\sim 10^7\ \mathrm{days\ cm^{-3}}$. This is also among the highest values of $\tilde{t}_{\mathrm{rel}}$ for any ion, and can therefore be used as an upper limit to $\tilde{t}_{\mathrm{rel}}$. For the set of parameters which we use in this paper ($A_{*}=V_{4}=1$, see Sect.~\ref{sect:param}) we find that this relation is fulfilled for both the solar and higher metallicity models up to several hundred days for all interesting values of $\eta $. Only for shocks with very high velocity and low mass loss rate is the expansion timescale shorter than the ionization timescale.

\begin{figure}[h]
\begin{center}
\resizebox{\hsize}{!}{\includegraphics{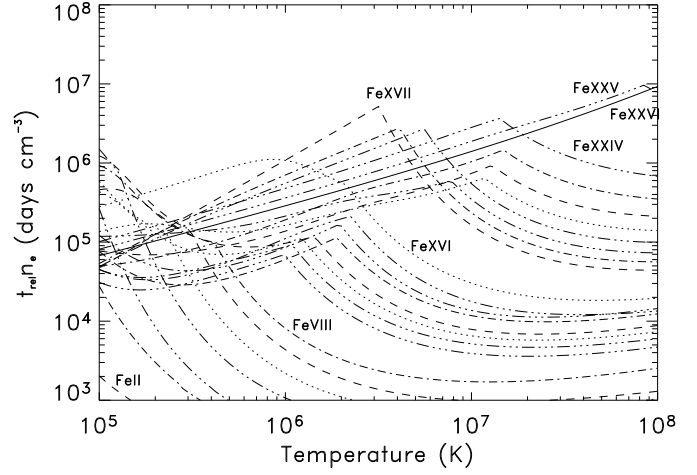}}
\end{center}
\caption{The lines represent the collisional ionization parameter, $\tilde{t}_{\mathrm{rel}}=t_{\mathrm{rel}}n_{\mathrm{e}}$ for Fe, with $t_{\mathrm{rel}}$ in days. } 
\label{fig:ionTime}
\end{figure}

For steady state ionization to hold, the relaxation time for ionization must also be shorter than the local cooling time, $\tilde{t}_{\mathrm{rel}}<t_{\mathrm{cool}}$. The cooling time decreases rapidly as the temperature drops in the post-shock gas, and a shock for which steady state holds at the shock front may be out of equilibrium further downstream. In terms of  the collisional ionization parameter and the cooling function we get

\begin{equation}
\label{eq:limitIonPar}
 \tilde{t}_{\mathrm{rel}}<\frac{3\mu kT}{\mu _{\mathrm{A}}\Lambda }. 
\end{equation}

\noindent
Using the first term of the cooling function in Eq.~(\ref{eq:coolFunc}), for $T_{e}\la 10^7$~K, we can express the condition for steady state ionization as

\begin{equation}
\label{eq:limitIonPar2}
\tilde{t}_{\mathrm{rel}}<\frac{4.8\times 10^{-15}\mu }{\mu _{\mathrm{A}}A}T_{6}^{1+\alpha }\ \mathrm{days\ cm^{-3}}.
\end{equation}

\noindent
This requirement should be checked for each ion at the temperatures where the ion is abundant. Note that this only depends on the temperature and composition, but not on the density.

As an example we consider Fe in the solar composition model, for
which the collisional ionization parameter is shown in
Fig.~\ref{fig:ionTime}. Using $\tilde{t}_{\mathrm{rel}}\sim 10^7\
\mathrm{days\ cm^{-3}}$ as an upper limit to
$\tilde{t}_{\mathrm{rel}}$, as discussed above, we can determine a
lower limit to the temperature, $\sim 5\times 10^6$~K, above which
steady state can be assumed to hold. Below this temperature time
dependent effects are expected to become important, although the fact
that $\tilde{t}_{\mathrm{rel}}$ varies with the temperature, and that
the value used here is an upper limit, means that steady state will be
a reasonable approximation also at somewhat lower temperatures. The
corresponding requirement for the helium model is $T\ga 1.8\times
10^6$~K, while for the carbon and oxygen dominated models $T\ga 1.1\times 10^7$~K and  $T\ga 3\times 10^7$~K, respectively. 

From this we expect steady state to be a reasonable approximation for the X-ray emitting part of a solar composition and a helium dominated shock. The carbon and oxygen dominated shocks, on the other hand, are expected to be out of equilibrium already at the shock front. For these shocks a fully time dependent ionization balance calculation is required, as we indeed do in this paper.

To illustrate this quantitatively we show in
Figs.~\ref{fig:ionStructFe-steady-timeDep} and
\ref{fig:ionStructFe-steady-timeDep-O} the ionization structure of Fe
both for steady state and time-dependent ionization in our standard
model with solar composition and oxygen dominated composition,
respectively. We see that in the solar case the state of ionization
behind the shock follows steady state  until the gas has cooled to a
few times $10^5$~K. After that, recombination begins to lag behind
cooling, the effects being noticeable in \ion{Fe}{vi} and lower ions,
which is what we would expect from the estimate above. Since this occurs at such low temperatures, it has no
noticeable effect on the X-ray emission. 

For the carbon and oxygen dominated models non-equilibrium effects are important already for \ion{Fe}{xxi} at $7\times 10^6$~K, as we expected. As extreme examples, we note in Fig.~\ref{fig:ionStructFe-steady-timeDep-O} that the concentrations of \ion{Fe}{xvii} and \ion{Fe}{xviii} stay high at considerably lower temperatures than in the steady-state case, due to the long recombination time compared to the cooling time. Except for $\lambda \ga 100$~\AA , the effects on the spectra are, however, minimal.

From this discussion we conclude that the equilibrium assumption is reasonable, except for low temperature shocks, and shocked gas highly enriched in heavy elements. This is useful since it allows a much simplified calculation of the shock emission, without a fully time dependent formalism.

\begin{figure}[h]
\begin{center}
\resizebox{\hsize}{!}{\includegraphics{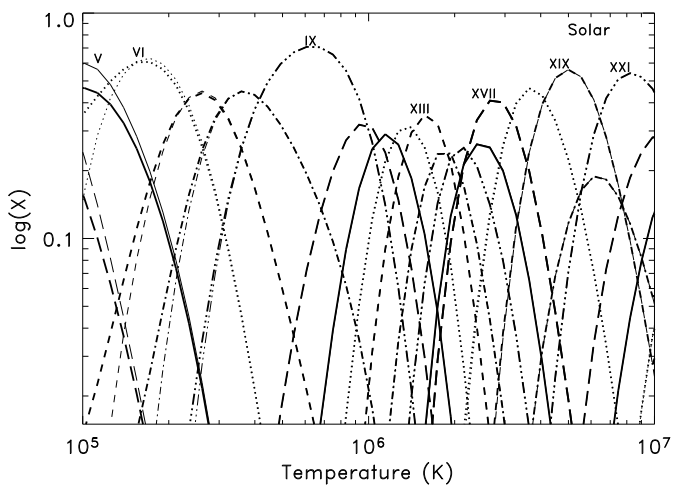}}
\end{center}
\caption{Ionization structure of Fe behind the reverse shock as function of temperature. The thin lines are the steady state model, and the thick lines are for a time-dependent ionization. The input parameters are the same as for the model in Figs.~\ref{fig:tempDens} and~\ref{fig:iondensFe}, and the composition is solar.} 
\label{fig:ionStructFe-steady-timeDep}
\end{figure}

\begin{figure}[h]
\begin{center}
\resizebox{\hsize}{!}{\includegraphics{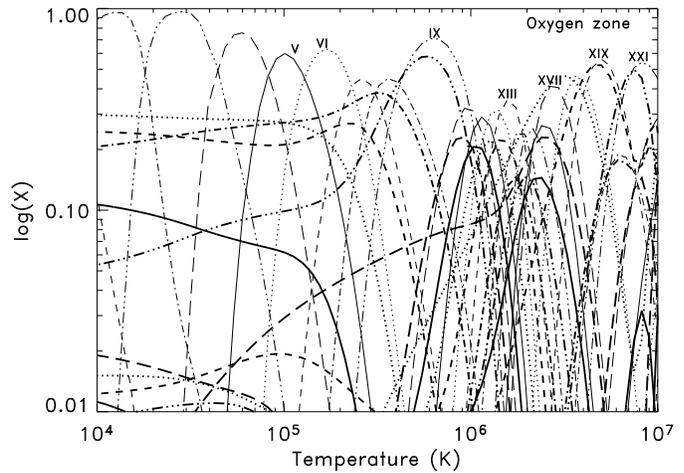}}
\end{center}
\caption{The same as in Fig.~\ref{fig:ionStructFe-steady-timeDep}, but for an oxygen dominated composition.} 
\label{fig:ionStructFe-steady-timeDep-O}
\end{figure}

\subsection{Optical depth and density effects}
\label{sect:optDepth}
Radiative transfer is taken into account by means of
the escape probability formalism, described in Sect.~\ref{sect:emission}, but turns out to have little effect. To understand the reason for this, we estimate the optical depth, $\tau $, for typical parameters. Combining Eqs.~(\ref{eq:optDepth}) and~(\ref{eq:crossSect}), and using the total size of the region, $D$, instead of the Sobolev length, as is more realistic,

\begin{equation}
\label{eq:optDepth2}
\tau \sim 4.9\times 10^{-12}n_{\mathrm{e}}DA(Z)X_{m}\frac{f_{kj}}{E_{kj}T^{1/2}},
\end{equation}

\noindent
where $A(Z)$ is the abundance of element $Z$, $X_{m}$ is the ionization
fraction of ion $m$, and the energy of the transition, $E_{kj}$, is in eV. 

As particularly interesting examples we consider two of the strongest
lines, the He-like \ion{O}{vii} $\lambda 21.60$ and the H-like \ion{O}{viii} $\lambda 18.97$
resonance lines, with $f=0.69$ and $f=0.27$, respectively. For collisional
ionization equilibrium the maximum abundance of \ion{O}{vii} occurs at $T
\sim 8\times 10^5$ K, where $X_m \sim 1$, and for \ion{O}{viii} at $T \sim
2.5\times 10^6$ K, where $X_m \sim 0.5$. If we now write
\begin{equation}
\label{eq:optDepth5}
\tau = \tau_0 A(Z) N_{21},
\end{equation}
where $N_{21}$ is the total column density in units of
$10^{21}~\mathrm{cm}^{-2}$, we find that $\tau_0 = 6.6\times 10^{3}$
for \ion{O}{vii}  $\lambda 21.60$, and $\tau_0 = 6.4\times 10^{2}$ for \ion{O}{viii} $\lambda 18.97$. 

The column density of the cooling gas is 

\begin{equation}
N \approx \frac{1}{4}t_{\mathrm{cool}} V_{\mathrm{rev}}
n_{\mathrm{i}} = \frac{3}{8} \frac{k T_0 V_{\mathrm{rev}}}{(1-\mu /\mu _{\mathrm{A}}) \Lambda(T_0)},
\end{equation}

\noindent
and therefore depends only on the composition and reverse shock
velocity, but not on the mass loss rate or shock radius. For solar
composition and $T_0 \sim 10^7$ K we get $N \sim 10^{21} \ \mathrm{cm^{-2}}$. Further, $A({\mathrm{O}}) = 6.9\times 10^{-4}$, and we therefore obtain
$\tau \sim 4.4$ for \ion{O}{vii} $\lambda 21.60$, and $\tau = 0.44$ for
\ion{O}{viii} $\lambda 18.97$.

For an oxygen dominated composition we get in the same way $\tau \sim
5.6\times 10^3 N_{21}$ for \ion{O}{vii} $\lambda 21.60$. Because of the stronger cooling the radiating region is also much
thinner (Sect.~\ref{ssect:shockStruct}). Our model calculation yields
a thickness of $4\times 10^{11}$~cm and a column density $N_{21}\sim 6.3\times 10^{-3}$ for the same input parameters as above. Therefore, the higher oxygen abundance results in an optical depth of \ion{O}{vii} that is comparable to that of the solar model. Some of the strongest lines may therefore be marginally optically thick. The great majority have, however, $\tau \la 1$.

A consequence of an optical depth higher than unity is that collisional
de-excitation or transfer between different L and S states may become
important. The ratio of collisional de-excitation to radiative
escape is given by $n_{\mathrm{e}} C_{kj}/ \beta_{kj} A_{kj} \approx n_{\mathrm{e}} C_{kj}
\tau_{kj}/ A_{kj}$, for $\tau_{kj} \ga 1$.

For the \ion{O}{vii} resonance line the most important destruction channels
are either by collisional de-excitation, or by a collisional
transition to the $2s\,^1S_{0}$ level, followed by a two-photon
transition.  At $10^6$ K the collisional de-excitation rate is $C_{2s\,^3P_{1} \rightarrow 1s\,^1S_{0}}=2.4\times10^{-10} \ \mathrm{cm^{3}} \
\mathrm{s}^{-1}$, while the radiative transition rate is $A_{2s\,^3P_{1}
\rightarrow 1s\,^1S_{0}}=3.2\times10^{12} \ \mathrm{s^{-1}}$. Therefore, there is a critical density, 
$n_{\mathrm{crit},2s\,^3P_{1}-1s\,^1S_{0}}= 2.2\EE{16} N_{21}^{-1} A(\mathrm O)^{-1}\cmc$, above which collisional destruction dominates radiative deexcitation.  The
forbidden $2s\,^3S_{1} \rightarrow 1s\,^1S_{0}$ line has $C_{2s\,^3S_{1} \rightarrow 1s\,^1S_{0}} = 4.1\EE{-11}\  \mathrm{cm^3 \ s^{-1}}$
and $A_{2s\,^3S_{1} \rightarrow 1s\,^1S_{0}} = 1.1\EE{3}\  \mathrm{s^{-1}}$, and $n_{\mathrm{crit},2s\,^3S_{1}-1s\,^1S_{0}}= 2.7\EE{13}\cmc$. Transitions
to the singlets occur with $C_{2s\,^3S_{1} \rightarrow 2s\,^1S_{0}}=1.2\times10^{-10} \ \mathrm{cm^{3} \ s^{-1}}$, so conversion to
two-photon emission occurs at a somewhat lower density, $\sim
10^{13} \cmc$. To summarize, conversion to
two-photon emission or destruction by collisional de-excitation will
only be important above $\sim 10^{13} \cmc$, independent of the
optical depth. 

Before this occurs, collisional excitation from $2s\,^3S_{1}$ to
$2p\,^3P_{2}$ will for 
$n \ga n_{\mathrm{crit},2s\,^3S_{1}-2p\,^3P_{2}} = 7\EE{10} \cmc$ convert radiative decays in the
$\wl 22.10$ \ $2s\,^3S_{0} \rightarrow 1s\,^1S_{0}$ line to the
$\wl 21.80$ \ $2p\,^3P_{2} \rightarrow 1s\,^1S_{0}$ line. Because
of limited instrumental resolution, as well as velocity smearing,
the individual lines will, however, be difficult to resolve. The total flux in
the $\wl 21.60 - 22.10$ \ lines will therefore not be affected below
$\sim 10^{13} \cmc$.

To check the dependence on the density and optical depth we have
made calculations for two sets of
parameters, one set with $\tilde{A}_{*}=1$ at 100 days, and one set
with $\tilde{A}_{*}=10$ at 10 days, and both with $\eta =10$, $s=2$ and $T_{0}=1$~keV. The density behind the shock in the latter
calculations is therefore a factor of $10^3$ higher than in the
`standard' calculations below. This range covers most cases of
interest.

After adjusting for the factor of ten difference in flux, because of
the difference in mass loss rates, we find that only the He-like \ion{C}{v} $\wl 41.47$ \ $2 {}^3S_{0} \rightarrow 1 {}^1S_{0}$ and $\wl 40.73$ \ $2 {}^3P_{2} \rightarrow 1 {}^1S_{0}$ lines, and the
corresponding \ion{O}{vii} $\wl 22.10$ \ and  $\wl 21.80$ \ lines are
affected by more than 50\%, as predicted from the simple arguments
above. Other lines are largely unaffected by the difference in density.

We therefore expect that neither density nor optical depths should affect the fluxes appreciably, except for some He-like
ions. Even in these cases the effect is likely to be hidden by the
instrumental resolution or velocity broadening.

\subsection{Variation with reverse shock velocity and composition}
From the discussion in Sect.~\ref{sect:timescales} we find that as long as the steady state requirement is fulfilled, the density behind the reverse shock is of secondary importance for the spectrum, except as a normalization parameter (Eq.~\ref{eq:lum}). We therefore now discuss the effects of varying the remaining parameters, the shock temperature and the composition.

\subsubsection{Reverse shock velocity}
\label{sect:velocityEffects}
The temperature of the post-shock gas depends on the velocity of the reverse shock.  In Fig.~\ref{fig:specComp3Trev} we compare the spectra from four models with solar composition, where the reverse shock temperature, $T_{0}$, is 0.3 keV, 1.0 keV, 3.0 keV and 10.0 keV, respectively. For a given shock velocity, each temperature corresponds to a particular value of the ejecta density gradient given by Eqs.~(\ref{eq:Trev}) and (\ref{eq:Tcs}). Here we have fixed the maximum velocity of the ejecta at $V_{4}=1$. The above reverse shock temperatures then correspond to density gradients  $\eta =22.0$ for $T_{0}=0.3$~keV, $\eta =8.3 $ for $T_{0}=3.0$~keV and $\eta =5.5 $ for $T_{0}=10.0$~keV. The ejecta velocity is only important for the velocity broadening of the lines. To show the distribution of the flux with wavelength more clearly, we plot the luminosity per logarithmic wavelength interval, $\lambda L_{\lambda}$.

\begin{figure*}[t]
\begin{center}
\includegraphics[width=17cm]{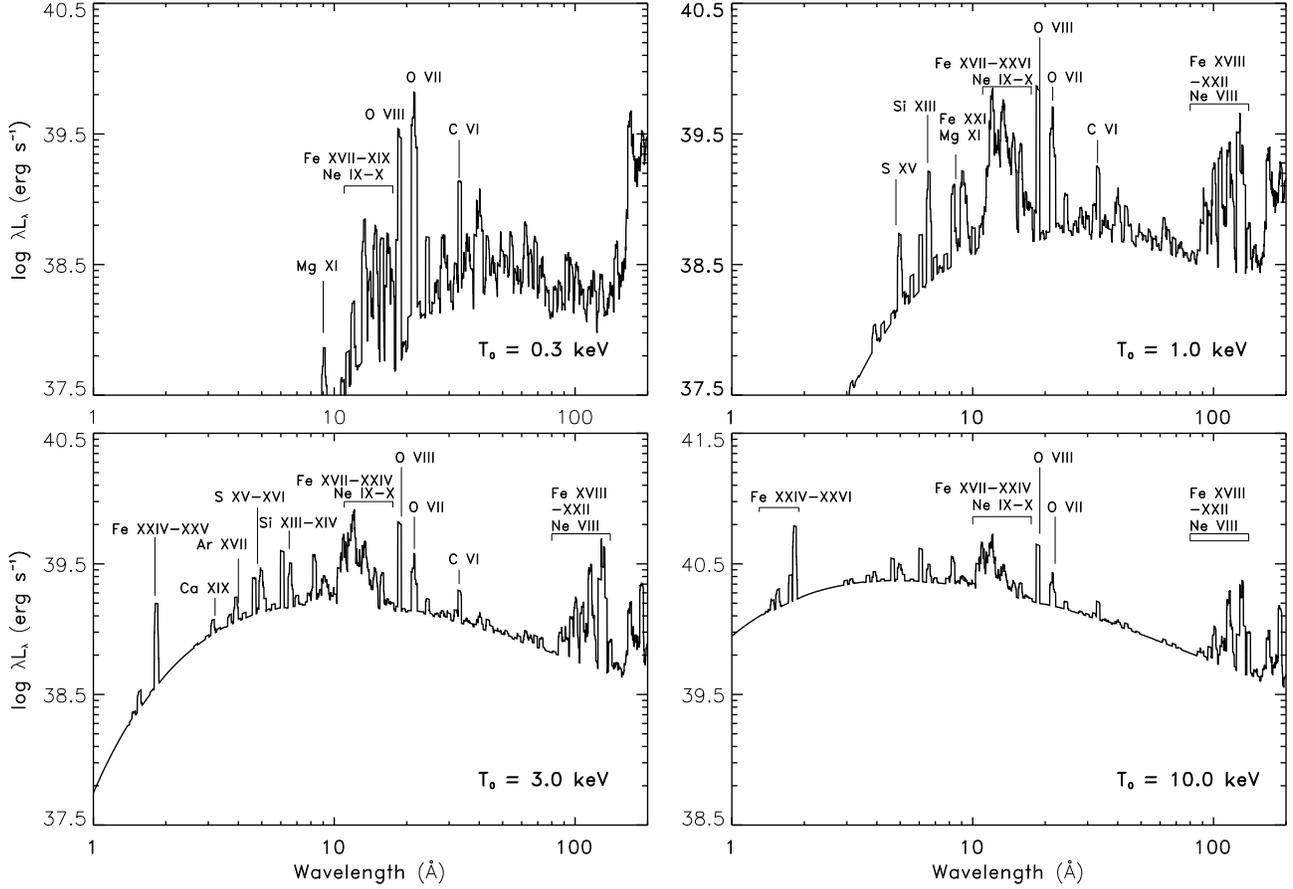}
\end{center}
\caption{Spectra with different reverse shock
temperatures, $T_{0}=0.3,\ 1.0,\ 3.0\ \mathrm{and}\ 10.0$~keV, all for solar composition.}
\label{fig:specComp3Trev}
\end{figure*}

The most obvious difference between the four models is the high energy exponential cutoff. Further, a higher temperature leads to a higher degree of ionization immediately behind the shock. However, because the gas is cooling, low and intermediate ionization stages are present even in the hottest model. In the models with $T_{0}\ga 3$~keV, the hot gas behind the shock produces little or no line emission, while contributing strongly to the continuum emission, which overwhelms the line emission from the cooler region. 

Fig.~\ref{fig:lineRatios} shows the ratio of the strongest lines to the flux of the strong \ion{O}{viii} $\wl 18.97$ line as a function of $T_{0}$ for solar composition. The left hand panel shows the strongest lines of the metals up to Ca, while the right hand panel shows the strongest Fe lines. Note that these relative luminosities do not include the continuum contribution. It is seen from this that at low temperatures the line emission is dominated by C, N and O, while the H- and He-like ions of Mg, Si and S become stronger for $T_{0}\ga 1$~keV. Lines of especially \ion{Fe}{xix}, \ion{Fe}{xxi}, \ion{Fe}{xxiii}, \ion{Fe}{xxiv} and \ion{Fe}{xxv} dominate successively the cooling for $T_{0}\ga 0.4$~keV.

The most conspicuous features in the $T_{0}=1$~keV and $T_{0}=3$~keV models are the Fe complexes between 10--15~\AA\ and around 100~\AA . The complex at 10--15~\AA\  consists of emission from \ion{Fe}{xxi--xxiv}, mixed with \ion{O}{viii}
 and \ion{Ne}{ix-x}, while the 100~\AA\ feature is a blend of lines from lower ionization stages of Fe, \ion{Fe}{xviii--xxii}, as well as \ion{O}{vii-viii}, \ion{Ne}{viii} and \ion{Ni}{xxiii--xxv}. These features are also present in the $T_{0}=10$~keV model, but have a smaller equivalent width due to the strong continuum. In the $T_{0}=0.3$~keV model the Fe features are practically absent, because the high ionization stages of Fe are no longer present. This can be seen in Fig.~\ref{fig:ionStructFe-steady-timeDep}, which shows that at a temperature of $\sim 3.5\times 10^6$~K \ion{Fe}{xviii} is the most abundant ion, while higher ionization stages of Fe are practically non-existent. Lower ionization stages of Fe have few emission lines at these energies, and the line emission in this model is therefore taken over by lighter metals. 

\begin{figure*}[t]
\begin{center}
\includegraphics[width=12cm,angle=-90]{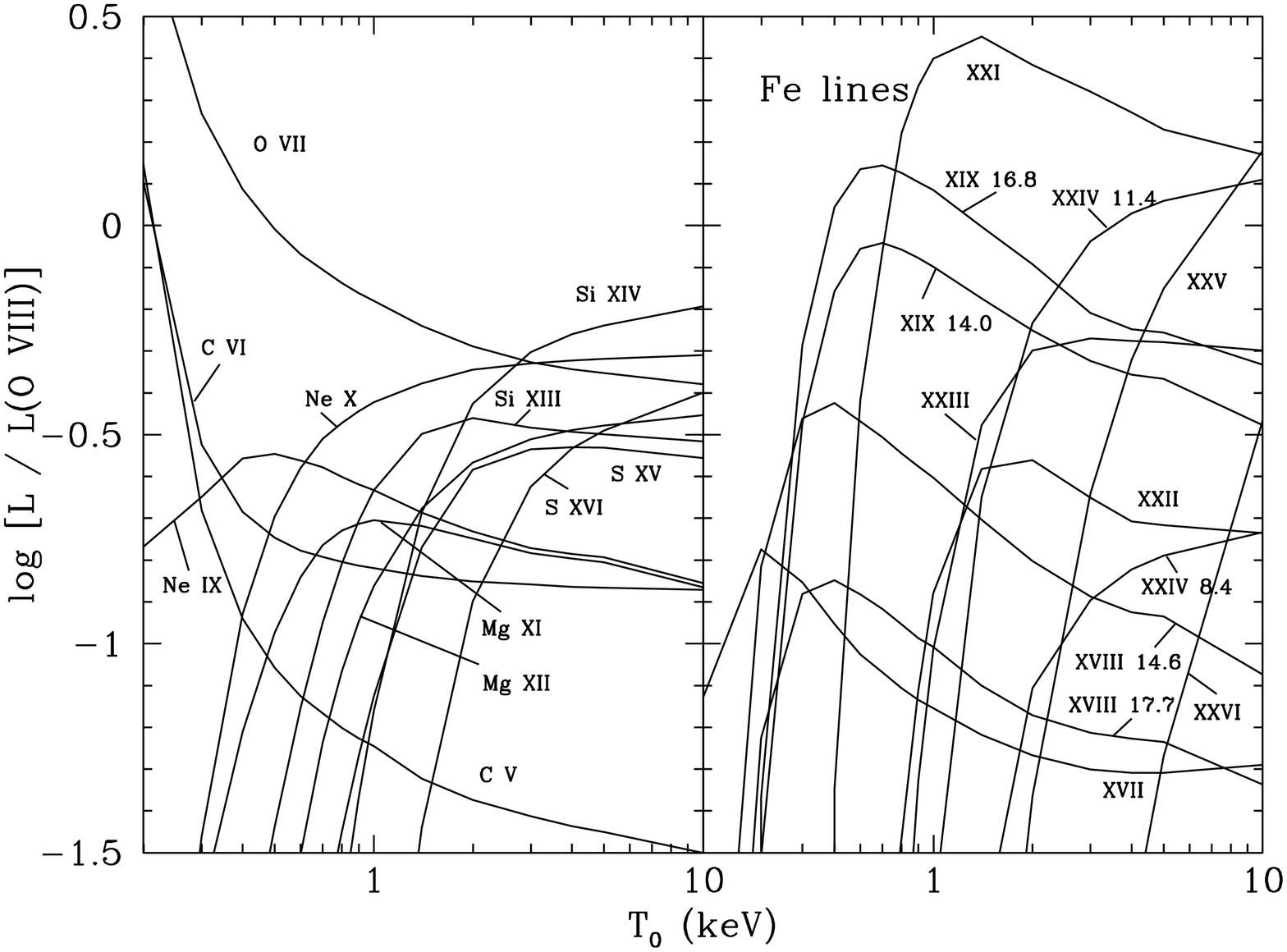}
\end{center}
\caption{Luminosities of the strongest lines relative to \ion{O}{viii} $\wl$18.97 as
function of the temperature of the reverse shock. The flux weighted
wavelengths of the different lines labeled are \ion{C}{v} $\wl$41.5, \ion{C}{vi} $\wl$33.7, \ion{O}{vii} $\wl$22.2, \ion{Ne}{ix} $\wl$13.7, \ion{Ne}{x} $\wl$12.1, \ion{Mg}{xi} $\wl$9.3, \ion{Mg}{xii} $\wl$8.4, \ion{Si}{xiii} $\wl$6.7,
\ion{Si}{xiv} $\wl$6.2, \ion{S}{xv} $\wl$5.1, \ion{S}{xvi} $\wl$4.7, \ion{Fe}{xvii} $\wl$17.1, \ion{Fe}{xviii} $\wl\wl$14,6, 17.7, \ion{Fe}{xix}
$\wl\wl$14.0, 16.8, \ion{Fe}{xx} $\wl$15.5, \ion{Fe}{xxi} $\wl$14.2, \ion{Fe}{xxii} $\wl$13.2, \ion{Fe}{xxiii} $\wl$12.2, \ion{Fe}{xxiv} $\wl\wl$8.4, 11.4,
\ion{Fe}{xxv} $\wl$1.87, \ion{Fe}{xxvi} $\wl$1.78.
}
\label{fig:lineRatios}
\end{figure*}

Another line complex is seen at 175--200~\AA , which consists mainly of lines from \ion{Fe}{x-xi}, with contributions from intermediate stages of O, Ne, Ar, Ca and Ni. This feature is strong in the $T_{0}=0.3$~keV model, but weakens with increasing shock temperature. 

The emission at $\lambda \la 10$~\AA\ comes from H-like and He-like stages of Ne, Mg, Si, S, Ar, Ca, Fe and Ni. From Fig.~\ref{fig:lineRatios} it is apparent that these lines become increasingly stronger as the temperature increases.  In the $T_{0}=3$~keV spectrum the Fe~K emission is the strongest, with the \ion{Fe}{xxv} 1.85~\AA\ line dominating, as well as weaker \ion{Fe}{xxv} $\wl\wl$ 1.46, 1.57-1.58 lines.  In the $T_{0}=10$~keV model the Fe~K lines at 1.4-1.6~\AA\ are strong, and now include \ion{Fe}{xxv-xxvi}, as well as a contribution from \ion{Ni}{xxvii} at 1.60~\AA .

In the $T_{0}=0.3$~keV model we also note the \ion{O}{viii} absorption edge at 17 \AA , which is not seen in the other models.

\subsubsection{Effects of composition}
\label{sect:compEffects}
As described in Sect.~\ref{sect:composition}, we have calculated four models with different chemical composition, one with solar composition, and three corresponding to the most important burning zones in the progenitor (Table~\ref{tab:abund}). In Fig.~\ref{fig:shockModAbundSpec} the spectra of the four models are shown. In all cases $T_{0}=1$~keV. As we explained in Sect.~\ref{sect:param}, we assume that $V_{4}=1$ in all cases, and to get the same $T_{0}$ for all compositions we have varied $\eta $. For $V_{4}=1$ a reverse  shock temperature $T_{0}=1.0$~keV corresponds to $\eta =13.0$ for the solar model, $\eta =18.3 $ for the helium dominated model, $\eta =20.2 $ for the carbon dominated model and $\eta =20.7$ for the oxygen dominated model.

The most conspicuous features in the helium and solar models are the Fe line complexes around 10~\AA\ and 100~\AA . These features are enhanced by lines from \ion{Ne}{ix-x} near 10~\AA , and \ion{Ne}{vii-viii} and \ion{O}{vii} near 100~\AA , otherwise Fe dominates these features. The higher abundance of C, O and Ne in the helium model means that these elements dominate the cooling, which can also be seen from the strength of the lines, particularly from \ion{C}{v-vi} and \ion{Ne}{ix-x}. Therefore, lines from heavier metals are stronger in the solar model than in the helium model, in spite of their lower abundance. This is especially apparent below 10~\AA , where the line emission in both models comes from H-like and He-like Mg, Si and S and  He-like Ar, all of which are stronger in the solar model than in the helium model.

In the carbon and oxygen dominated models the Fe features are more or less absent. This is because of the strong overabundance of O and Mg, as well as Si and S, which take over the cooling in the high energy part of the spectrum. Therefore, although the Fe abundance is higher than in the other models, the Fe lines are overwhelmed by the stronger emission from lighter elements. Consequently, instead of the Fe features near 10 \AA\ and 100~\AA , the oxygen model has strong lines from a number of ions distributed over the whole spectrum, while the carbon model has few, but strong lines.

The line emission of the carbon model is dominated by lines from \ion{C}{v-vi}, \ion{O}{vii-viii} and \ion{Ne}{vii-x}. The oxygen model, on the other hand has only weak C emission and no Ne lines at all as a result of the low C and Ne abundances. Instead, it  is dominated by emission from \ion{O}{vii-viii}, mixed with strong lines from H-like and He-like Mg, Si and S, and lower ionization stages of the same elements between 10 -- 100~\AA .

The continuum of the carbon and oxygen models is considerably steeper at long wavelengths compared to the solar and helium models. This reflects the increasing importance of bound-free and two-photon contributions compared to the free-free emission in the metal rich models. As a result both the carbon and oxygen models show strong absorption edges, which are not apparent in the solar model. The most prominent is the \ion{O}{ix} edge at 13.8~\AA , which is also present, albeit weak, in the helium model. In addition, we see strong edges from \ion{Ne}{xi} at 51.7~\AA , and \ion{C}{vii} at 24.8~\AA\  in the carbon model, and from  \ion{Mg}{xiii} at 7.3~\AA\ and \ion{Mg}{xiv} at 6.5~\AA\ in the oxygen model. Also the absorption edges of \ion{Ne}{viii} and \ion{Si}{vii}, which coincide at 62.0~\AA , are apparent in the oxygen spectrum. In both the carbon and oxygen models the signature of two-photon emission from \ion{O}{vii} and \ion{O}{viii} is apparent, most clearly in the carbon dominated model, where it is not obscured by line emission.

\begin{figure*}[t]
\begin{center}
\includegraphics[width=17cm]{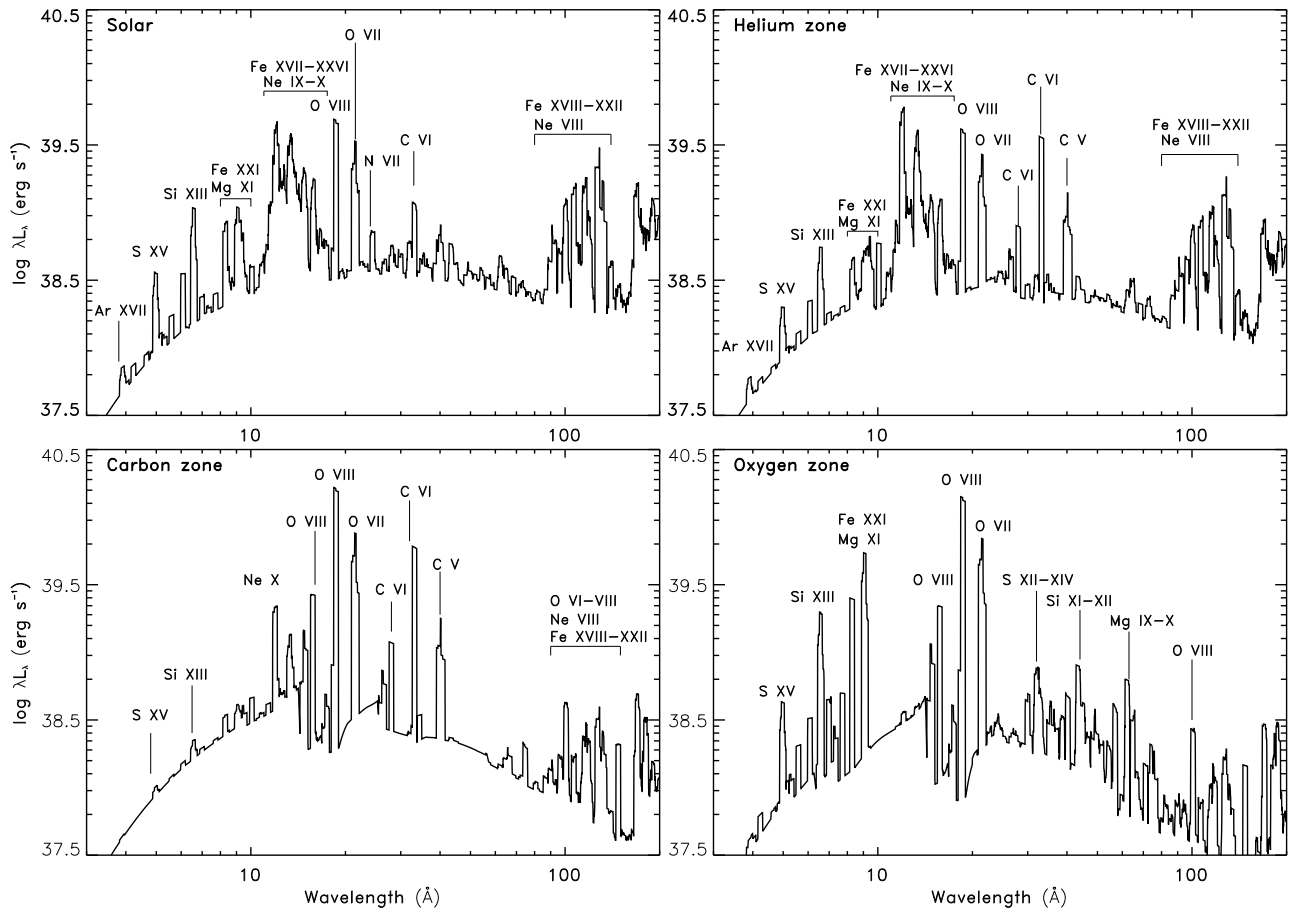}
\end{center}
\caption{X-ray spectra produced by four models with different composition. All models have $V_{4}=1$ and $T_0=1.0$~keV. The chemical abundances in each zone are listed in
Table~\ref{tab:abund}.}
\label{fig:shockModAbundSpec}
\end{figure*}

\subsection{Effects of pre-ionization}
As mentioned previously the state of ionization immediately behind the shock is determined by that of the pre-shock ejecta gas. For the 1~keV case most pre-shock oxygen is in \ion{O}{v-vii}, while iron is found as \ion{Fe}{viii-xi}, both for solar and oxygen dominated composition. Only after $\sim \tilde{t}_{\mathrm{rel}}$ does the ionization reach the state corresponding to the local temperature. 

In Fig.~\ref{fig:preIon} we show the difference between the emitted spectrum when pre-ionization is included, and when we assume that ionization equilibrium is reached instantly, for a solar composition and for an oxygen dominated gas. In the solar mixture the difference is barely noticeable, except for very short wavelengths, and even there it is negligible. In the oxygen dominated gas, on the other hand, the difference is pronounced for wavelengths shorter than $\sim 15$~\AA . The reason for this is the much shorter cooling time in the oxygen dominated model compared to the solar model. The steepening continuum emission at short wavelengths in the oxygen dominated model is caused by the dominance of free-bound and two-photon emission over free-free emission, since  both free-bound and two-photon processes are strongly affected by the lower ionization state in the post-shock gas when pre-ionization is included. For the line emission, the hydrogen-like lines are most affected by pre-ionization, as expected. We note e.g.,  that the hydrogen-like \ion{Mg}{xii} line at 8.4~\AA\ is a factor $\sim 3$ lower when pre-ionization is included. The helium-like \ion{Mg}{xi} line at 9.3~\AA , however, is hardly affected at all. This is as expected, since the effect of pre-ionization is to reduce the abundance of high ionization stages close to the shock front.

\begin{figure*}[t]
\begin{center}
\includegraphics[width=17cm]{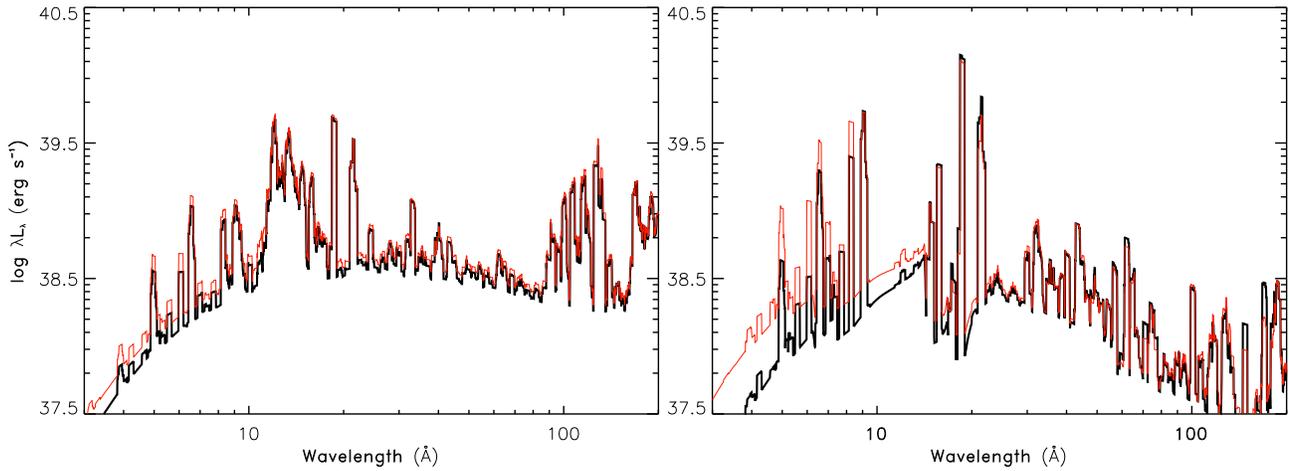}
\end{center}
\caption{Comparison between spectra where pre-ionization is included (thick line) and spectra where ionization equilibrium is assumed to be reached instantly in the post-shock gas (thin line). The left hand panel is for solar composition and the right hand panel is for an oxygen dominated composition. In all cases $V_{4}=1$ and $T_0=1.0$~keV.}
\label{fig:preIon}
\end{figure*}

\subsection{Comparison with single-temperature models}
\label{sect:comp1T}
When X-ray observations of supernovae are analyzed, the emitting region is usually
assumed to have a single temperature. In some cases two temperatures
are used, to account for the different temperatures of the two
shocks. As an example, Uno et al. (\cite{Uno02}) fit the ASCA observations of
SN~1993J with two components, with different temperatures and
absorption column densities. However, while a single temperature fit
is a good approximation for the 
outer shock, which is adiabatic, it is far from sufficient for reproducing the
emission from a radiative shock, as we will show below. 

Another interesting example is the Chandra observations of SN~1993J after 8 years, which could
only be fit with two low temperature
components (0.35 keV and 1.01 keV) and one high temperature component
(6.0 keV) (Swartz et al.~\cite{Swartz03}). Similarly Schlegel et al. (\cite{Schlegel04}) find, when trying to fit
spectral models to recent Chandra observations of SN~1978K, that while
a two component model gives the best fit of the models tested, the
fit even for this model was not satisfactory. This indicates that more
detailed models are necessary in order to fit the observed spectra. In
particular, it is important to realize that  the relative contribution
of the different temperature components can not be treated as free parameters, but are fixed by the hydrodynamic structure of the shock as it cools.

In Fig.~\ref{fig:specComp1THydro} a spectrum produced by our combined hydro-/spectral code, with $T_{0}=1.0$~keV, is compared to a single temperature spectrum with a temperature equal to the post-shock temperature in the whole region between the reverse shock and contact discontinuity. The spectra are normalized to have the same total luminosity between 1 -- 10 keV. Some of the most prominent lines in each model are given in Table~\ref{tab:lineStrengths}, where the fluxes relative to one of the most prominent lines in the spectra, \ion{O}{viii} $\lambda $~18.97, are listed for these two models, as well as for a single temperature model with $T=0.7$~keV. 

As expected, the continuum emission in the cooling
model is higher at low energies, and there are more lines from lower ionization
stages  than in the single-temperature model. In particular, we see that the line
complexes around 10~\AA\ and 100~\AA\ are broader in the full model
than the single-temperature model, a result of the wider range of ionization stages contributing to these features. At energies below 10~\AA , the line
emission in both models is dominated by H-like and He-like Mg, Si and S, the H-like lines being stronger in the single-temperature model than in
the full model. At wavelengths longer than $\sim 15$~\AA\ the line emission is stronger
in the full model than in the single-temperature model, due to the
presence of colder gas. The lines in the single-temperature model
mainly come from \ion{Fe}{xx--xxiv}, together with weaker lines from \ion{C}{vi}, \ion{O}{viii} and 
\ion{Ne}{x}, while the high temperature does not
allow lower ionization stages. In the cooling model the same lines are
present, but also lines of lower ionization stages like \ion{C}{v}, \ion{N}{vii} and \ion{O}{vii}, as well as
\ion{Si}{ix--xiii}, \ion{S}{xi--xiv}, \ion{Ar}{xiv--xvi}, \ion{Ca}{xv--xviii} and \ion{Fe}{xv--xvi}. 

Quantitatively, the difference can be seen from Table~\ref{tab:lineStrengths}. The line ratios of, e.g., the H-like lines of the different elements differ by large factors between the two models. In reality one would not use the temperature behind the shock for a single-temperature model, but instead try to find the temperature which gives the best fit to the observed spectrum. Because of the contributions from cooler zones, the best-fit spectrum will correspond to a lower temperature than the shock temperature. When attempting to fit the shock model with $T_{0}=1.0$~keV shown in Fig.~\ref{fig:specComp1THydro} with a single-temperature model, we find the best overall fit to be a single-temperature model with $T=0.7$~keV.  Although there is a slight improvement compared to the single-temperature $T=1$~keV model, the line ratios differ by at least a factor of two from the shock model. In addition, the 'best-fit' temperature is too low. To reproduce the line emission
from both high and low ionization stages a
combination of cold and hot gas is therefore necessary. Note, however, that a multi-component model
is not sufficient, because \emph{the contributions from each of these
components have to be calculated self-consistently, as in the shock
model.} The large differences between these models are of obvious importance when determining abundances, and a too simplified analysis can lead to large errors in the results.

\begin{table*}[h*]
\caption{Line strengths relative to the \ion{O}{viii} $\wl 18.97$ line for a shock model with $T_0=1.0$~keV, a single temperature model at the same temperature and a single temperature model at the temperature, 0.7~keV, which gives the best fit to the spectrum of the shock model.}
\label{tab:lineStrengths}
\begin{center}
\begin{tabular}{lcccc} \hline \hline 
Ion & Wavelength (\AA )& F/F(\ion{O}{viii})  & F/F(\ion{O}{viii}) & F/F(\ion{O}{viii}) \ \\ 
& & Shock model & Single temperature (1 keV) & Single temperature (0.7 keV)\ \\ \hline \\
\ion{C}{vi} & 33.73 & 0.15 & 0.10 & 0.10 \\
\ion{N}{vii} & 24.78 & 0.07 & 0.05 & 0.05 \\
\ion{O}{vii} & 21.60 & 0.23 & 0.02 & 0.03 \\
\ion{Ne}{x} & 12.13 & 0.38 & 0.72 & 0.65 \\
\ion{Mg}{xii} & \phantom{ } 8.42 & 0.14 & 0.60 & 0.29\\
\ion{Si}{xiii} & \phantom{ } 6.67 & 0.16 & 0.65 & 0.34\\
\ion{S}{xv} & \phantom{ } 5.00 & 0.05 & 0.35 & 0.09\\
\ion{Fe}{xxi} & 12.29 & 0.46 & 1.60 & 1.15 \\
\hline
\end{tabular}
\end{center}
\end{table*}

\begin{figure}[h]
\begin{center}
\resizebox{\hsize}{!}{\includegraphics{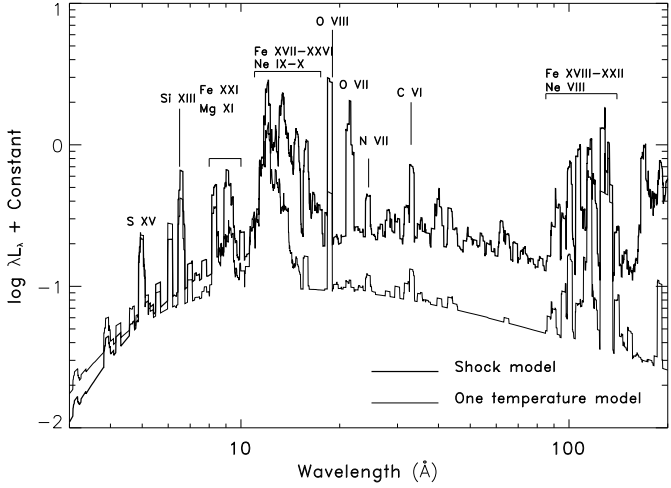}}
\end{center}
\caption{Comparison between a single temperature spectrum with $T=1.0$~keV (thin line) and a cooling shock spectrum created by our model, with $T_0=1.0$~keV (thick line), both with solar composition. The spectra are normalized to have the same total flux in the range 1--10~keV.}
\label{fig:specComp1THydro}
\end{figure}

\subsection{Absorption by the cool shell}
\label{ssect:absorption}
When the reverse shock is radiative, most of the emission from the reverse shock is absorbed by the cool shell close to the contact discontinuity. At early epochs, only the most energetic emission passes through, while lower
energies are completely absorbed.  As the emitting region expands, the
column density in the cool shell drops, and the shell becomes transparent, first
at high energies, later at lower energies. 
The absorption can be taken into account by calculating the emerging luminosity at wavelength $\lambda $ as

\begin{equation}
 L_{\lambda }=L_{\lambda,\mathrm{em}}\mbox{e}^{-\tau (\lambda)} \ ,
 \end{equation}

\noindent
where the subscript `em' refers to the emitted luminosity, $\tau =N_{\mathrm{col}}\sigma (\lambda)$, and $\sigma (\lambda)$ is the absorption cross section at wavelength $\lambda $, given by 

\begin{equation}
\label{eq:crossSect2}
\sigma (\lambda )=\sum _{m,Z}X_{m}A(Z)\sigma _{m,Z}(\lambda)\ .
\end{equation}

This obviously depends on the composition, but is relatively insensitive to the state of ionization of the cool shell, as long as K-shell opacity dominates. For solar composition $\sigma (\lambda )\approx 2.2\times 10^{-25}\lambda ^{8/3}\ \mathrm{cm}^{2}$~(\cite{CF94}), where the wavelength $\lambda $ is given in \AA . The column density of cool gas along the line of sight depends on the swept up mass behind the reverse shock according to Eqs.~(\ref{eq:MswCS}) and (\ref{eq:MswRev}), so that 

\begin{eqnarray}
\label{eq:colDens}
N_{\mathrm{cool}}&\approx &\frac{M_0}{4\pi R^2\mu m_{\mathrm{u}}}=\frac{(\eta -4)}{8\pi}\frac{\dot M}{v_{\mathrm{w}}\mu _{\mathrm{A}}m_{\mathrm{u}}}\frac{1}{r} \nonumber \\
&=&1.8\times 10^{23}\frac{(\eta -4)}{\mu _{\mathrm{A}}}\tilde{A_{*}}V_{4}^{-1}t_{\mathrm{d}}^{-1}\ .
\end{eqnarray}

Clumping of the cool, dense shell  may, however, influence the column
density~(Chevalier \& Blondin~\cite{ChevBlond95}). From their
hydrodynamic calculations Chevalier \& Blondin find a variation of
$N_{\mathrm{cool}}$ by a factor $\sim  10$ in different
directions. The total emission will then be a superposition of
emission along lines of sight with different column densities. Since
we generally do not know the degree of clumpiness, we, here, choose to
treat the column density as a free parameter, adjusting the absorption
to fit the observations. For illustrative purposes we show in
Fig.~\ref{fig:specAbs} the effects of absorption using three different
values for the column density, $10^{21} - 10^{23}\ \rm{cm}^{-2} $. It is obvious that the low energy
cutoff is very sensitive to this parameter, and for a detailed
modeling a superposition of several absorption components, as given by
the hydrodynamic calculations, may be needed. 

\begin{figure}[h]
\begin{center}
\resizebox{\hsize}{!}{\includegraphics{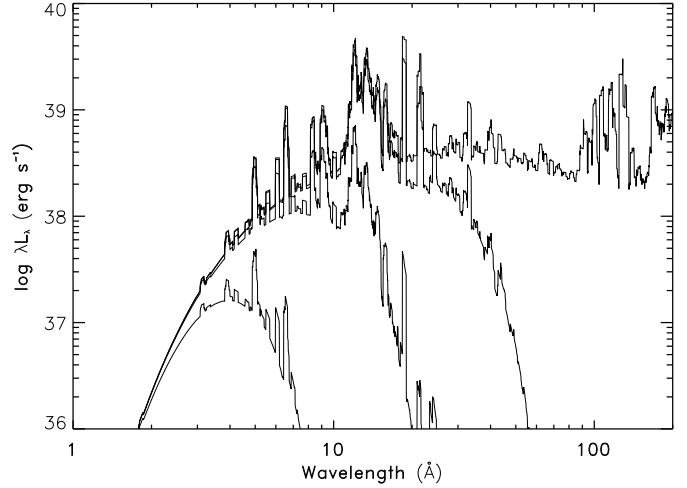}}
\end{center}
\caption{The uppermost spectrum shows the energy emitted by the reverse shock for the same model as in Fig.~\ref{fig:specComp1THydro}, while the lower three lines represent the emerging luminosity after absorption with column densities of $N_{\mathrm{cool}}=10^{21}\ \mathrm{cm}^{-2},\ 10^{22}\ \mathrm{cm}^{-2}\ \mathrm{and}\ 10^{23}\ \mathrm{cm}^{-2}$, respectively.} 
\label{fig:specAbs}
\end{figure}

\subsection{X-rays from the circumstellar shock}
\label{sect:csshock}
While this paper mainly concerns a detailed modeling of the X-rays
from the reverse shock, there is also a contribution from the outer,
circumstellar shock. Because of the high velocity, and consequently
high shock temperature, the emission from this will be
dominated by continuum emission. This has been discussed in detail in
\cite{FLC96}. A complication is here that the coupling between
electrons and protons by Coulomb collisions decreases with increasing
temperature, 

\begin{eqnarray}
t_{\mathrm{e-i}}&\approx &29 \left(\frac{T_{\mathrm{e}}}{10^{9}\
\mathrm{K}}\right)^{3/2}\left(\frac{n_{\mathrm{e}}}{10^8\
\mathrm{cm^{-3}}}\right)^{-1}\nonumber \\
& \approx & 0.21\left(\frac{T_{\mathrm{e}}}{10^{9}\
\mathrm{K}}\right)^{3/2}\tilde{A_{*}}^{-1}V_{4}^{2}t_{\mathrm{d}}^{2}\ \ \rm{days}.
\end{eqnarray} 

The electron temperature may, therefore,
be considerably lower than the ion temperature. In FLC96 it was found
that for the parameters of SN 1993J with a shock velocity of $\sim
2\times 10^4\ \kms$ and $\tilde{A}_{*}=5$ the ion temperature was
$(1-2)\times 10^{10}$~K, while the electron temperature was only $\sim
1\times 10^9$~K at 10 days. As the density is decreasing with
decreasing $\tilde{A}_{*}$ and increasing time, the difference between
the ion and electron temperatures can be even larger. A caveat here
is, however, that plasma instabilities may lead to more efficient
equipartition. The observations of SN 1993J, however, indicated that
this was not very important (FLC96). For the reverse shock the density is so high and the temperature so low that the electrons and ions will generally be in equipartition.

The spectrum and luminosity of the circumstellar shock depend on the
mass loss rate, $\tilde{A}_{*}$, shock velocity, $V_{\mathrm{s}}$, and
time since explosion, $t$. In contrast to the reverse shock, the
circumstellar shock is in general adiabatic. The temperature behind
the shock is therefore fairly uniform (see e.g., Fig. 5 in FLC96). To 
estimate the monochromatic luminosity in the X-ray band we use a
similar method to that in \cite{FLC96}, with $s=2$. Approximating the free-free Gaunt
factor, $g_{\mathrm{ff}}$, (including relativistic effects and both
ion-electron and electron-electron emission) with
$g_{\mathrm{ff}}\approx 2.3T_{9}^{0.42}\lambda ^{0.23}$ for $1\la
\lambda \la 10$~\AA , we get

\begin{eqnarray}
\label{eq:Lcs}
L_{\lambda,\mathrm{cs}}& \approx &9.6\times 10^{37}\zeta \tilde{A}_{*}^{2}V_{4}^{-1}\left (\frac{f}{0.2}\right )^{-1} \nonumber \\
& &\times T_9^{-0.08}\lambda ^{-1.77} e^{-0.144/\lambda T_9} \nonumber \\
& & \times \left
(\frac{t}{100\  {\mathrm{days}}}\right )^{-1}\ \ \mathrm{erg\
s^{-1}\AA ^{-1}}.
\end{eqnarray}

\noindent
Here $T_{9}=T_{\mathrm{e}}/10^9$~K, $\lambda $ is in \AA ,  $\zeta
=[1+2n(\mathrm{He})/n(\mathrm{H})] /
[1+4n(\mathrm{He})/n(\mathrm{H})]$ and $f$ is the relative thickness of the
circumstellar shock region, $f =(R_{\mathrm{cs}}-R_{\mathrm{s}})/R_{\mathrm{s}}$. For $\eta =7$ we have $f\approx 0.3$, and
for $\eta \ga 12$, $f\approx 0.22$ (\cite{CF94}). The luminosity from Eq.~(\ref{eq:Lcs}) should be added to the flux from the reverse shock, reduced by the factor below, in Eq.~(\ref{eq:occultation}). 

The ejecta are in general optically thick, and absorb the emission from that part of the interaction region which is occulted by the ejecta. For the reverse shock, where the cooling region is thin, the difference in radius is so small that we can approximate the observed luminosity as one half of the emitted luminosity. This has been taken into account when plotting the model spectra. For the outer shock, on the other hand, the size of the shocked region can be as much as $20-30\%$ of the total radius, and this approximation is inaccurate. The relation between the observed and emitted luminosity can then be expressed as

\begin{equation}
\label{eq:occultation}
\frac{L_{\lambda ,\mathrm{obs}}}{L_{\lambda ,\mathrm{em}}}=0.5+0.5\frac{((f+1) ^2 -1)^{3/2}}{(f +1)^3 -1}
\end{equation} 

For $f =0.25$ this gives $L_{\mathrm{obs}}=0.72L_{\mathrm{em}}$.  Thus, in Fig.~\ref{fig:totSpec} we have reduced the outer shock emission by this factor, while the reverse shock emission has been reduced by one half, as in all other spectra.

In addition to the free-free emission, inverse Compton scattering of
the photospheric photons by the thermal, semi-relativistic electrons
behind the circumstellar shock may contribute to the X-ray emission
(Fransson \cite{Fransson1982}, Lundqvist \& Fransson \cite{Lundq88},
FLC96). The flux of this depends sensitively on the temperature of the
electrons, the mass loss rate, and the optical and UV flux of the
supernova. In general, this component is only important for electron
temperatures $\ga 10^9$ K, and during the first months after explosion, 
and then only for high mass loss rates. This was discussed in some
detail for SN 1993J (FLC96), where it was concluded that because
of the slow equipartition between electrons and ions, the electron
temperature was too low to scatter the photospheric photons up to the
X-ray regime. A characteristic signature of a Compton component would
be a powerlaw component to the X-rays.

Also inverse Compton by the relativistic electrons, responsible for the radio emission, may contribute to the X-rays. This component is, however, only likely to be important for low values of $\tilde{A}_{*}$ (Chevalier et al.~\cite{CFN05}).

As an example we show in Fig.~\ref{fig:totSpec}  a full spectrum including the thermal emission from both shocks, and the absorption by the shell, for $\tilde{A}_{*}=1$,
$T_{0}=1$,~keV, $V_{4}=1$, $\eta =13.0$ and $ N_{\mathrm{cool}}=10^{22}\
\mathrm{cm^{-2}}$ at 100 days. At this epoch the ion temperature
behind the circumstellar shock is $\sim 2\times 10^9$ K, but the
electron temperature is only $\sim (1-4)\times 10^8$ K, due to the
inefficient Coulomb coupling at the low density, $\sim 2\times 10^6 \ \rm cm^{-3}$, behind the
circumstellar shock. The cutoff in
the hard spectrum is therefore at $\sim 0.4$ \AA, or $\sim 30$ keV. 

For  $Ê\lambda \la 3$~\AA\ (E $\ga 3$~keV) the spectrum is dominated by the circumstellar shock. This is also the case above the cutoff wavelength from the cool, dense shell at $\sim 20$~\AA . The figure illustrates the importance of taking both the absorption of the cool, dense shell and the circumstellar component into account when analyzing the spectrum. Note also that the column density of the cool shell is quite uncertain (Sect.~\ref{ssect:absorption}). In addition, absorption from the interstellar medium, and possibly also the circumstellar medium, should be included.

\begin{figure}[h]
\begin{center}
\resizebox{\hsize}{!}{\includegraphics{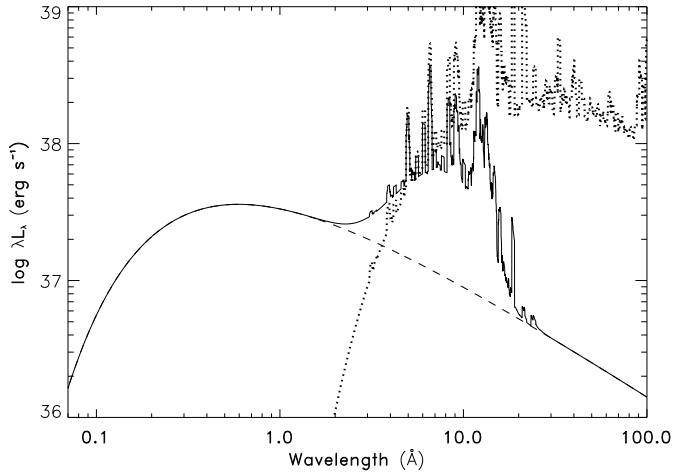}}
\end{center}
\caption{Total spectrum for $V_{4}=1$, $\tilde{A}_{*}=1$, $\eta =13.0$, $T_{0}=1$~keV, $t=100$~days and solar composition. The solid line is the sum of the emission from the circumstellar and reverse shocks, assuming an absorbing column density of $N_{\mathrm{cool}}=10^{22} \ \mathrm{cm^{-2}}$ in the cool shell. The dotted line is the unabsorbed contribution from the reverse shock, and the dashed line is the contribution from the circumstellar shock. } 
\label{fig:totSpec}
\end{figure}

\section{Conclusions}
\label{sect:concl}
This paper discusses for the first time a detailed, self-consistent model for the X-ray emission from supernovae interacting with a dense circumstellar medium. The model which we have presented here is intended to fill
a gap in the modelling of these supernovae, and is applicable to the
case where the reverse shock is radiative. Our model differs from
earlier models in that the spectral code is coupled to a
hydrodynamical code, and that we solve the full multi-level
statistical equations of each ion to calculate the line emission. The
cooling and ionization, which are computed for each temperature and
density, are used in the hydrodynamical equations to determine the
size of the emitting region. Thus the total emission from each zone is
determined self-consistently, giving the correct weight to the
emission from each zone.

We have shown that for a sufficiently steep ejecta density gradient the radiative model can be applicable up to several years, depending on the progenitor's mass loss rate. We have also shown that its applicability depends strongly on the composition of the cooling gas, and that the high metallicity shocks stay  radiative much longer than a shock with a solar composition.

A stationary shock has been found to be a good approximation for the cases of interest to us, while the ionization must be considered in a time-dependent formulation to account for the non-equilibrium recombination in the cooling gas. This effect is, however, mainly important for lines with $\lambda \ga 100$~\AA . We have shown that the emerging spectrum is highly sensitive to both the temperature of the reverse shock and the composition of the gas. Because of the strong dependence of the emerging spectrum on the composition of the gas, the chemical abundances in the ejecta can be determined from the spectrum. 
A requirement for this is, however, a self-consistent modeling of the hydrodynamic and radiative processes, as demonstrated in this paper. 

\begin{acknowledgements}
We are grateful to Roger Chevalier and Peter Lundqvist for comments and a careful reading of the
manuscript, and to  Ken Nomoto for discussions. This work was supported in part by the Swedish Research Council and the Swedish National Space Board.
\end{acknowledgements}

\bibliographystyle{aa}

\appendix
\suppressfloats
\section{Atomic data}
\label{app}
The elements included in the model are H, He, C, N, O, Ne, Mg, Si, S, Ar, Ca, Fe and Ni. Table~\ref{tab:ionLev} lists the ions for which level population calculations are performed, along with the number of levels included for each ion.

\begin{table*}[p*]
\centering
\caption{Number of levels and transitions included in the model for each ion.}
\label{tab:ionLev}
\begin{tabular}{lclclclc} \hline \hline
Ion & Lev./Trans. & Ion & Lev./Trans. & Ion & Lev./Trans. & Ion & Lev./Trans. \\ 
\hline
\ion{H}{i} & 25/75 &  \ion{Mg}{v} & 10/25 & \ion{S}{xiv} & 40/117  & \ion{Fe}{x} & 172/3959 \\
\ion{He}{i} & 16/187 & \ion{Mg}{vi} & 23/142 & \ion{S}{xv} & 49/166  & \ion{Fe}{xi} & 47/168 \\
\ion{He}{ii} & 25/68 & \ion{Mg}{vii} & 46/389 & \ion{S}{xvi} & 25/68 & \ion{Fe}{xii} & 41/696 \\
\ion{C}{ii} & 18/40 &  \ion{Mg}{viii} & 125/654 & \ion{Ar}{iv} & 30/165 & \ion{Fe}{xiii} & 27/87\\
\ion{C}{iii} & 20/53 & \ion{Mg}{ix} & 46/313 &  \ion{Ar}{vii} & 16/35 & \ion{Fe}{xiv} & 40/227 \\
\ion{C}{iv} & 15/18 & \ion{Mg}{x} & 40/123 & \ion{Ar}{viii} & 21/34  & \ion{Fe}{xv} & 53/935 \\
\ion{C}{v} & 49/165 & \ion{Mg}{xi} & 49/166 & \ion{Ar}{ix} & 89/1175 &   \ion{Fe}{xvi} & 21/34 \\
\ion{C}{vi} & 25/68 & \ion{Mg}{xii} & 25/68 & \ion{Ar}{x} & 3/3 &  \ion{Fe}{xvii} & 89/463\\
\ion{N}{i} & 26/72 & \ion{Si}{ii} & 15/30 &  \ion{Ar}{xi} & 10/24 & \ion{Fe}{xviii} & 113/176\\
\ion{N}{ii} & 23/86 & \ion{Si}{iii} & 20/67 & \ion{Ar}{xii} & 72/1776 & \ion{Fe}{xix} & 92/1276\\
\ion{N}{iii} & 20/56 & \ion{Si}{iv} & 21/34  & \ion{Ar}{xiii} & 15/44 &  \ion{Fe}{xx} & 86/1309\\
\ion{N}{iv} & 20/59 & \ion{Si}{v} & 27/115 & \ion{Ar}{xiv} & 125/654 &  \ion{Fe}{xxi} & 290/6435\\
\ion{N}{v} &15/18 &  \ion{Si}{vi} & 3/3 & \ion{Ar}{xv} & 46/136  & \ion{Fe}{xxii} & 204/6820\\
\ion{N}{vi} & 49/173 & \ion{Si}{vii} &  10/25 &  \ion{Ar}{xvi} & 40/139 & \ion{Fe}{xxiii} & 58/258  \\
\ion{N}{vii} & 25/68 & \ion{Si}{viii} & 72/2061 & \ion{Ar}{xvii} & 49/166 &  \ion{Fe}{xxiv} & 40/155\\
\ion{O}{ii} & 15/48 & \ion{Si}{ix} & 46/403 & \ion{Ar}{xviii} & 25/68 & \ion{Fe}{xxv} & 49/168 \\
\ion{O}{iii} & 46/335 & \ion{Si}{x} & 125/654 & \ion{Ca}{ix} & 16/35 & \ion{Fe}{xxvi} & 25/68 \\
\ion{O}{iv} & 125/654 &  \ion{Si}{xi} & 46/134 & \ion{Ca}{x} & 21/34 & \ion{Ni}{xii} & 31/101\\
\ion{O}{v} & 20/57 &  \ion{Si}{xii} & 40/134 &  \ion{Ca}{xi} & 89/971 & \ion{Ni}{xiii} & 48/191\\
\ion{O}{vi} & 40/107 & \ion{Si}{xiii} & 49/166 & \ion{Ca}{xii} & 3/3 & \ion{Ni}{xv} & 27/87\\
\ion{O}{vii} & 49/173 & \ion{Si}{xiv} & 25/68 & \ion{Ca}{xiii} & 20/63 &  \ion{Ni}{xvi} & 40/246 \\
\ion{O}{viii} & 25/68 & \ion{S}{ii} & 43/284 & \ion{Ca}{xiv} & 15/96 & \ion{Ni}{xvii} & 16/35\\
\ion{Ne}{ii} & 3/3 & \ion{S}{iii} & 49/366 & \ion{Ca}{xv} & 46/421 & \ion{Ni}{xviii} & 21/34\\
\ion{Ne}{iii} & 58/1212 & \ion{S}{iv} & 52/220 & \ion{Ca}{xvi} & 125/632 & \ion{Ni}{xix} & 89/336\\
\ion{Ne}{iv} & 22/91 &  \ion{S}{v} & 16/35 & \ion{Ca}{xvii} & 46/136 &  \ion{Ni}{xx} & 113/176\\
\ion{Ne}{v} & 49/408 & \ion{S}{vi} & 21/34 & \ion{Ca}{xviii} & 40/136 & \ion{Ni}{xxi} & 58/1265  \\
\ion{Ne}{vi} & 125/550 & \ion{S}{vii} & 27/111 & \ion{Ca}{xix} & 49/166 &  \ion{Ni}{xxii} & 15/96\\
\ion{Ne}{vii} & 46/127 & \ion{S}{viii} & 3/3 &  \ion{Ca}{xx} & 25/68 & \ion{Ni}{xxiii} & 20/79\\
\ion{Ne}{viii} & 40/103 & \ion{S}{ix} &  86/2640 & \ion{Fe}{ii} & 142/1268 &  \ion{Ni}{xxiv} & 125/654 \\
\ion{Ne}{ix} & 49/173 & \ion{S}{x} & 22/215 & \ion{Fe}{vi} & 80/2083  &  \ion{Ni}{xxv} & 46/136\\
\ion{Ne}{x} & 25/68 & \ion{S}{xi} & 46/416 & \ion{Fe}{vii} & 9/23 & \ion{Ni}{xxvi} & 40/161 \\
\ion{Mg}{ii} & 21/34 & \ion{S}{xii} & 125/654  & \ion{Fe}{viii} & 83/267 &  \ion{Ni}{xxvii} & 49/168\\
\ion{Mg}{iv} & 3/3 & \ion{S}{xiii} & 46/136 & \ion{Fe}{ix} & 13/45  & \ion{Ni}{xxviii} & 25/68 \\
\hline
\end{tabular}
\end{table*}

The ionization rates (collisional and autoionization) are computed according to 
Arnaud \& Rothenflug~(\cite{ArnRoth}), except for Fe, where we use
the method suggested by Arnaud \& Raymond~(\cite{ArnRay}). 

The recombination rates are evaluated as the sum of the
radiative and dielectronic recombination coefficients, $\alpha = \alpha
_r + \alpha _d$. For the radiative recombination rates of C, N and O we use the fits
tabulated by Nahar~(\cite{Nah96}, \cite{Nah99}) and Nahar \&
Pradhan~(\cite{NahPr97}), except for C-like N and O (see below). For C-like, O-like, F-like and Ne-like ions the recombination rates are evaluated from the fitting parameters of Zatsarinny et al.  (\cite{Zat03}, \cite{Zat04a}, \cite{Zat04b}), using  the fitting formula suggested by Verner \& Ferland~(\cite{VF96}): 

\begin{eqnarray}
\label{eq:ionfitVF}
\alpha _r(T) & = & a[(T/T_0)^{1/2}(1+(T/T_0)^{1/2})^{1-b} \nonumber \\
& \times &
(1+(T/T1)^{1/2})^{1+b}]^{-1}. 
\end{eqnarray}

\noindent
The same formula is used for H-like through Ne-like ions of Mg, Si, S, Ar, Ca, Fe, and Ni (Gu~\cite{Gu03}) and for  H-like, He-like, Li-like and Na-like ions
of the remaining elements (Verner \& Ferland~\cite{VF96}). For all remaining ions we use the rates compiled by D. Verner \footnote{Recombination rates are available from http://www.pa.uky.edu/$\sim $verner/fortran.html.}, which include data from Aldrovandi \& Pequignot~(\cite{AlPeq73}), Shull \& Van Steenberg~(\cite{SvS82}), Arnaud \& Raymond~(\cite{ArnRay}), Landini \& Monsignori Fossi~(\cite{LM90}, \cite{LM91}), Pequignot et al.~(\cite{Peq91}) and Verner \& Ferland~(\cite{VF96}).

The dielectronic recombination rates are fitted with 

\begin{equation}
\label{eq:fitDielRec}
\alpha _{d}(T)=\frac{1}{T^{3/2}}\sum _{i=1}^{4}c_{i}\mathrm{e}^{-E_{i}/kT}.
\end{equation}

\noindent
For C-like, O-like, F-like and Ne-like ions we use the fitting parameters of Zatsarinny et al.  (\cite{Zat03}, \cite{Zat04a}, \cite{Zat04b}) \footnote{Dielectronic recombination rates are available from http://homepages.wmich.edu/$\sim  $gorczyca/drdata/.}, and for the remaining ions we use the data from Mazzotta et al. (\cite{Mazzotta98}).

The collisional and radiative rates for the excited levels are taken from version 4.2 of the Chianti database (Dere et
al.~\cite{Dere97}; Young et al.~\cite{Young03})~\footnote{CHIANTI is a collaborative project involving the NRL (USA), RAL (UK), and the Universities of Florence (Italy) and Cambridge (UK). }, except for the recombination to individual levels. We include radiative recombination to He-like, Li-like and Na-like ions, using data from Verner \& Ferland~(\cite{VF96}), and dielectronic recombination to \ion{Fe}{xvi} and \ion{Fe}{xvii}, with data from Dasgupta~(\cite{Dasg95}).

\end{document}